\documentclass[12pt,preprint]{aastex}

\usepackage{graphicx}
\usepackage{fullpage,amsmath,amssymb,latexsym}
\usepackage{multirow}
\usepackage{rotating} 
\usepackage{natbib}
\usepackage{url}
\shorttitle{The Formation of Uranus \&Neptune}
\shortauthors{Helled \& Bodenheimer}

\begin{document}

\title{The Formation of Uranus \& Neptune: \\ Challenges and Implications For Intermediate-Mass Exoplanets}
\author{Ravit Helled$^1$ and Peter Bodenheimer$^2$}
\affil{$^1$Department of Geophysical, Atmospheric, and Planetary
Sciences, Raymond and Beverly Sackler Faculty of Exact Sciences, 
Tel Aviv University, Tel Aviv, Israel \\
$^2$UCO/Lick Observatory, Department of Astronomy and Astrophysics, 
University of California, Santa Cruz,  CA 95064, USA}

\begin{abstract}
In this paper we investigate the formation of Uranus and Neptune, according to the 
core-nucleated accretion model, considering  
formation locations ranging from 12 to 30 AU from the Sun, and with various 
disk solid-surface densities and core accretion rates. It is shown 
that in order to form Uranus-like and Neptune-like planets in terms of final mass 
{\it and} solid-to-gas ratio, very specific conditions are required. We also show that 
when recently proposed high solid accretion rates are assumed,  along with 
solid surface densities about 10 times those in the minimum-mass
solar nebula, the challenge in forming Uranus and 
Neptune at large radial distances is no longer the formation timescale, but is rather 
finding agreement with  the 
final mass and composition of these planets. In fact, these conditions are more likely to
lead to gas-giant planets. 
Scattering of planetesimals by the forming planetary core is found to be an
important effect at the larger distances.  Our study emphasizes how (even 
slightly) different conditions in the protoplanetary disk and the birth environment 
of the planetary embryos can lead to the formation of very different planets in 
terms of final masses and compositions (solid-to-gas ratios), which naturally explains 
the large diversity of intermediate-mass exoplanets. 
\end{abstract}


\section{Introduction}
The increasing number of detected exoplanets with masses similar to those of Uranus 
and Neptune emphasizes the need to better understand the formation process of the
 planetary class consisting of planets with rock-ice cores of up to  
$\approx$ 15 M$_\oplus$
and lower-mass hydrogen/helium envelopes,  with a large range of 
solid-to-gas ratios. 
However, the formation mechanism for 
intermediate-mass planets is not well understood (e.g., Rogers et al. 2011), and
even within our own solar system there are many open questions regarding the 
formation of Uranus and Neptune. \par

Uranus and Neptune have masses of about 14.5 and 17 M$_{\oplus}$, and are located at 
19.2 and 30 AU, respectively. Their exact compositions are not known (e.g., Helled 
et al. 2011), but they are likely to consist mainly of rock and ices with smaller mass
fractions of hydrogen-helium atmospheres (see Fortney \& Nettelmann 2010; Nettelmann 
et al. 2013 and references therein). The estimated solid-to-gas mass 
ratios of Uranus and Neptune range between 
2.3 and 19 (Guillot 2005; Helled  
et al. 2011). Hereafter when we refer to the solid-to-gas ratio it 
should be clear that the solid component consists of all elements heavier 
than hydrogen and helium, and that in fact, their physical state can 
differ from a solid state, while the gas corresponds solely 
to hydrogen and helium. 
It is commonly assumed that Uranus and Neptune have formed by the core accretion 
scenario, in which solid core formation accompanied by slow gas accretion is followed by more
rapid gas accretion (e.g., Pollack et al. 1996). During the initial phases of planet 
formation, the slow accretion of gas  is controlled by the growth rate of its core. 
As a result, the solid accretion rate essentially determines the formation timescale 
of  an intermediate-mass planet (see D'Angelo et al. 2011 for a review). The (standard) solid accretion rate 
is given by (Safronov  1969):  
\begin{equation}
 \dot M_\mathrm{core} = {\frac{dM_\mathrm{solid}}{dt}}  = 
 \pi R_\mathrm{capt}^2
 \sigma_s \Omega  F_g,
\label{eq:accrete}
\end{equation} 
where $\pi R_\mathrm{capt}^2$ is the capture cross section for 
planetesimals, $\Omega$ is the orbital 
frequency, $\sigma_s$
is the solid surface density in the disk, and $F_g$ is the gravitational
enhancement factor. As can be seen from Equation~(1), the core accretion rate decreases
with increasing radial distance, and as a result, the core formation timescale can be 
extremely long at radial distances larger than $\sim$ 10 AU. Using the model of the 
minimum mass solar nebula (MMSN, see Weidenschilling 1977 for details), the formation 
timescales for Uranus and Neptune exceed $10^9$ years (Safronov 1969). This estimate  has 
introduced the {\it formation timescale problem of Uranus and Neptune}. \par

A detailed investigation of the formation of Uranus has been presented in Pollack et al.~(1996). 
The authors have considered {\it in situ} formation and have shown that for 
$\sigma_s$ about twice that of the MMSN the approximate core mass and envelope 
mass of Uranus are reached in about 16 Myr.  The timescale is considerably shorter
than that of Safronov primarily because of the use of  improved (and much higher) values 
for $F_g$. The core accretion rate, however, depends not only on $\sigma_s$ and $F_g$ but also
on  the  sizes of the accreted planetesimals. Smaller planetesimals can be accreted more 
easily, and it was shown by Pollack et al.~(1996) that when planetesimals are assumed 
to have sizes of 1 km (instead of the standard 100 km) the formation timescale of Uranus 
decreases to  $\approx$ 2 Myr. This timescale is within the estimated range 
of lifetimes of gaseous  protoplanetary disks but has been considered  unrealistically 
short because of the use  of a simplified core accretion rate in the model. 
 Clearly, the formation timescale for Neptune under similar assumptions 
would be significantly longer. Due to the long accretion times at large radial distances, 
Uranus and Neptune are often referred to as ``failed giant planets",  because their 
formation process was too slow to reach runaway gas accretion before the 
disk gas had dissipated. \par

Goldreich et al. (2004) discuss in detail the problem of the accumulation of solid
particles, particularly at large distances from the central star. Without considering the
effects of the gas, they suggest that to form Uranus- or Neptune-like 
planets {\it in situ},
one requires, first, relatively small planetesimals, $< 1$ km in radius, and second,
a value of $\sigma_s$ a few times that of the MMSN.
In fact, they conclude that in order to form the planets within the lifetime of the
gas disk, particles of only a few cm in size are needed. The small particles would be
generated by collisions between the km-size objects. Numerous collisions among
the small particles strongly damp the particle random velocities, resulting in a
very cold disk. However,  Levison \& Morbidelli (2007), on the basis of N-body
simulations,  point out that this model is oversimplified and relies on numerous
assumptions. They suggest that  the main difficulty is the assumption that the
surface density of the disk particles remains smooth and uniform. In fact the simulations
show that the formation of rings and gaps actually dominates the dynamics. 

The idea that the solar system was originally much more compact and that Uranus 
and Neptune were formed at smaller radial distances 
has been considered by a number of authors (e.g., Thommes et al. 
1999; Tsiganis et al. 2005). 
The planets must arrive at their present locations post-formation,  by gravitational scattering or by
migration induced by a disk of planetesimals. 
The success of the ``Nice Model'' to explain many of the observed properties of the solar system 
led Dodson-Robinson \& Bodenheimer (2010) to investigate the formation of 
Uranus and Neptune at radial distances of 12 and 15 AU, 
as suggested by that  model. They  adopted a 
disk model which accounts for disk evolution and disk chemistry (Dodson-Robinson et al. 2009), 
giving values of $\sigma_s$ at those distances an order of magnitude higher than in the MMSN.
The planet-formation calculation was similar to that of Pollack et al.~(1996).
It was found that the formation timescales of both 
Uranus and Neptune fell in the range 4--6 Myr, and that in some cases the solid-to-gas ratio was similar to those in the present planets. In addition, the results are consistent
with the observed carbon enhancement in the  atmospheres of these planets.
It was therefore concluded that, 
indeed, Uranus and Neptune could have formed at smaller radial distances as implied by 
the Nice model. There are, however, unsolved issues with this 
formation scenario---
for example, there is no way to distinguish Uranus from Neptune, and more importantly, 
there must be a cutoff of both solid and gas accretion when the planets reach their
current masses; otherwise the model predicts that they would continue to accrete to higher mass.
 \par 

While the scenario in which Uranus and Neptune form at smaller radial distances 
is feasible and somewhat promising, there is no evidence for ruling out the 
possibility that these planets, and in particular extrasolar planets,
could have formed at larger distances. Recent studies on 
the accretion rate of solids have provided new estimates for the rates in 
which solids are accreted to a planetary embryo in the core accretion paradigm 
(Rafikov 2011; Lambrechts \& Johansen 2012). These studies suggest that the 
solid accretion rates can be significantly higher than previous estimates. With 
these high accretion rates the core formation timescale at large radial distances 
is significantly reduced. Although {\it in situ} formation of Neptune is
considered unlikely (see below),  {\it in situ} formation of Uranus 
as well as formation of extrasolar giant planets, or Neptune-sized planets,
by core accretion at large radial distances might be feasible. \par 

The aim of this paper is to re-investigate the formation of Uranus and Neptune 
(as representatives of intermediate-mass planets) and, in particular,  to investigate the
consequences of employing the high accretion rates mentioned in the previous
paragraph. We account 
for various accretion rates, orbital locations, and disk properties. 
We employ a full core accretion--gas capture model. 
As we discuss below, formation  of planets at 20 and 30 AU is, in principle, feasible,
 although their characteristics may not be those of Uranus and Neptune. 
We also suggest that a major challenge in forming Uranus and Neptune is to derive the 
correct final masses and the solid-to-gas ratios. However, the sensitivity of the 
properties of the forming planets to the assumed parameters provides a natural 
explanation for the diversity of extrasolar planets in the Uranus/Neptune mass regime. \par

\section{The Formation Model and Its Parameters}
To model the formation of the planets we use a standard core accretion model 
(e.g., Dodson-Robinson \& Bodenheimer 2010; Lissauer et al. 2009), which combines
a given core accretion rate with a detailed model of the structure of the
gaseous envelope and for the interaction of incoming planetesimals with this
envelope. The planetary 
formation calculation  in core accretion theory usually
begins with a small seed body  
surrounded by a swarm of planetesimals. Clearly, the formation of the initial
core, even if small in mass, is not instant, 
 and in some cases the formation time  of the seed core can even be 
comparable to the planetary formation timescale. As a result, the formation 
timescales derived from  most of our simulations, especially when considering 
large radial distances and/or low solid-surface densities, should be taken as lower bounds. 
In the calculations presented here,
the initial core mass is about 1 M$_{\oplus}$, at which point a low-mass
gaseous envelope has been accreted. These planetary embryos are assumed 
to form at given radial distances, and migration during the formation process 
is neglected. In some of the cases we do account for disk evolution, by gradually
  reducing the gas (and therefore also the small-dust) surface density with time. 
The planetesimals are assumed to be of a 
single size, which is a free parameter in the model. As discussed below, the 
planetesimal's size has some  impact on the core accretion rates, but other
parameters are found to be more important. 

The structure of the gaseous envelope is calculated according to the
standard spherically symmetric equations of stellar structure. The
radiative opacities for the dust in the envelope are assumed to be reduced by 
approximately a factor 50 relative to standard interstellar dust
opacities (e.g.~Pollack et al. 1994). This reduction roughly
accounts for the settling and coagulation of dust grains in the envelope
(Movshovitz et al. 2010). The energy source in the envelope is
provided primarily by the accretion of planetesimals, although
gravitational contraction is also included. The gas accretion rate into
the envelope is determined by the requirement that the outer radius
of the planet ($R_\mathrm{pl}$) match the modified accretion radius
(Lissauer et al. 2009): $R_\mathrm{pl}^{-1} = R_B^{-1} + 4 R_H^{-1}$
where $R_H$ is the Hill radius (see below) and $R_B = GM_\mathrm{pl}/c_s^2$,
the Bondi radius.
Here $M_\mathrm{pl}$ is the total mass of the planet, and $c_s$ is
the sound speed in the disk. 

\subsection{Core Accretion Rates}
The formation of the planet is essentially determined by the growth rate 
of the core. However, the growth rates for solid cores in the solar 
nebula are unknown, and must be inferred from models. Not surprisingly, 
the estimates for the core accretion rates have a large range, which can lead 
to rather different outcomes in terms of planetary formation. 
In order to investigate the sensitivity of the planetary formation to the assumed 
core accretion rate we consider three different accretion rates. The first 
corresponds to a dynamically cold planetesimal disk, providing high accretion 
rates, and is given by (Rafikov 2011):  
\begin{eqnarray}
 \dot M_\mathrm{core}& \approx& 6.47 \Omega p^{1/2}\sigma_s R_H^2 \nonumber \\
&\approx& 8.5 \times 10^{-4} \sigma_s M_\mathrm{core}^{2/3}
\end{eqnarray}
where $M_\mathrm{core}$ is the core mass,  $\Omega$ is the orbital frequency, 
$p=R_\mathrm{core}/R_H$, and 
$R_H=a(M_\mathrm{pl}/(3 M_\odot))^{1/3}$, the Hill radius, where $a$ is
the distance of the planet from the star. The second expression applies 
if $M_\mathrm{pl} \approx M_\mathrm{core}$
and under our standard assumption that the core is a sphere of constant density  $\rho_\mathrm{core}=2$ g cm$^{-3}$. 
This accretion rate can be 
taken as the maximum accretion rate of solids according to  Rafikov (2011). It should be noted 
that in this model the sizes of planetesimals are small, since large planetesimals 
are assumed to fragment into small pieces that are later affected by gas drag leading 
to the state of a dynamically cold planetesimal disk. Therefore, when this 
accretion rate is considered, the calculation is more consistent when the planetesimals 
are assumed to be small ($<<$ 100 km). \par  

An alternative accretion rate was recently presented by Lambrechts \& Johansen (2012) 
who find that accretion of pebbles (cm-sized particles) 
 within the Hill radius is given by, 
\begin{equation}
 \dot M_\mathrm{core} = 2 R_H \sigma_s v_H, 
\end{equation}
where  $v_H$ is the relative velocity between 
the pebbles and the core, given by $v_H=\Omega R_H$. This expression applies
when the core mass is higher than the transition mass (their Eq. 33)
\begin{equation}
M_t \approx 3 \times 10^{-3} (\Delta/0.05)^3(a/5 AU)^{0.75} M_\oplus
\end{equation}
where  $\Delta= \Delta u_\phi/c_s$ represents the
difference between the mean orbital gas flow and a pure Keplerian orbit ($c_s$ is the disk
sound velocity). At 15 AU $M_t \approx 0.055$ M$_\oplus$; thus this accretion rate is valid
for all computations reported here.  These accretion rates are 
higher by about one order of magnitude than the maximum accretion rates 
derived from Equation~(2). In our work, we will set the high accretion rate 
to the one given by Equation~(2) unless differently stated, and refer to it as [dM/dt]$_{HIGH}$.

Finally, the lower-bound accretion rate used in our model is the one 
derived for the "transition case" (Rafikov 2011) in which 
\begin{equation}
\dot M_\mathrm{core} \sim (6-10)p R_H \sigma_s v_H,
\end{equation} 
where $p<<1$.
This rate refers to a ``warm'' planetesimal disk,  with  planetesimal random 
velocity dispersion $s_p$ on the borderline between
intermediate and high values ($s_p \approx \Omega R_H$).   
In the actual simulations we set the numerical coefficient equal to $2 \pi$.
The rate is similar to the one used by Dodson-Robinson \& Bodenheimer (2010) and 
is slightly lower than the accretion 
rates used by Pollack et al.~(1996). Therefore, the 
rate derived from Equation~(5) can be considered as  the low (or 
standard) core accretion rate; we refer to it as [dM/dt]$_{LOW}$. 
Even lower rates are of course possible. 
\par

It should be noted, however, that core accretion rates are hard to estimate,
and therefore the uncertainty in this quantity is rather large. 
As previously mentioned, detailed N-body simulations (Levison \& Morbidelli 2007) suggest 
that the collisional damping scenario, needed to produce a planetesimal disk with
low velocity dispersion, could be unreliable. Further N-body simulations (Levison et al. 2010)
show that the gravitational interactions between the embryo and the planetesimals
lead to the wholesale redistribution of material and to the opening of gaps near the embryos. 
If the region near the growing embryo is cleared of planetesimals before much growth 
can occur, the core accretion rate will decrease dramatically and will prevent the formation 
of a giant and/or intermediate mass planet. 
Nevertheless, the possibility of a high solid accretion rate cannot be excluded.
 Various authors suggest that high core accretion rates can be maintained also over 
long timescales due to fragmentation, even at 
large radial distance (Goldreich et al. 2004; Rafikov 2011 and references therein).  
In view of the uncertainties, we take the approach of selecting representative accretion rates, 
which cover a rather wide range. Thus  we can quantify the effect of the assumed core 
accretion rate (as well as other model assumptions) on the planetary growth. 

\subsection{Formation Locations}
There is no simple way to constrain the original radial distances at 
which 
planets form. This is due to the fact that planets are likely to change 
their positions due to interactions with the disk (e.g., Kley \& Nelson 2012) or with
other planets (Thommes et al.~1999).
Because of the location of the methane condensation front in the primitive
solar nebula (Dodson-Robinson et al.~2009), Uranus and Neptune most probably formed
beyond the orbit of Saturn, but whether they formed at the distances of $\sim$ 12-20 AU, 
as suggested by the Nice model, is yet to be determined.  
Formation of Neptune {\it in situ}
at 30 AU is generally considered unlikely, and Malhotra (1995) and 
Hahn \& Malhotra (1999) have shown that
many properties of the Oort cloud and the Kuiper belt can be explained  by the outward
migration of Neptune, driven by a planetesimal disk. Nevertheless we present
one test calculation involving formation at that distance. For our main
results, 
we consider three formation locations of the planets: 12, 15 and 20 AU, 
radial distances that are predicted by and are consistent with the formation of Uranus 
and Neptune in the Nice Model. It should be noted, however, that within 
the framework of that  model other orbital locations are possible as long 
as a separation of at least 2 AU between Uranus and Neptune is considered.  

\subsection{Solid Surface Densities}
A critical unknown property in planet formation models is $\sigma_s$, the 
solid surface density in the disk. This parameter is important because 
it determines how much material is available for the formation of a core 
at the location where  the planet is formed. In this work we consider 
various possibilities for the solid-surface density. In most of the cases we 
consider, the lower bound for $\sigma_s$ is similar to the one derived from the 
MMSM model (Weidenschilling 1977). The upper bound is typically about one order 
of magnitude higher than the MMSN model and is consistent with the solid 
surface densities derived by the disk model of Dodson-Robinson et al.~(2009), in 
which condensation of various species of ices as well as disk evolution are included. 
Other solid-surface densities are considered as well. 
%
The $\sigma_s$ that are used at various radial distances are listed in Table 1.

\subsection{Planetesimal Sizes and Enhanced  Planetesimal Capture Cross Section}
As already discussed in Pollack et al.~(1996), the planetary formation timescale 
decreases significantly when the accreted solids are small. Similarly to other 
disk properties, the sizes of planetesimals in the Solar Nebula are unknown, 
and in fact, they are likely to vary with radial distance and time. In addition, the 
planetesimals are expected to have a distribution of sizes. However, for simplicity, 
we consider single sizes for planetesimals. To investigate the effect of the 
planetesimal size on the planetary growth, we consider both small (1 km) and large 
(100 km) planetesimals. However the accretion rates given by Equations (2) and (3)
imply that planetesimal sizes are smaller than 1 km. Therefore we have run one
test with 50 meter planetesimals to investigate the effect of decreasing the size.
Our experiments with a range of over three orders of magnitude in planetesimal size make 
it clear what the effect would be if even smaller sizes were taken, as discussed below. 
\par 

The accretion rates defined by Equations (2) and (5) depend on the core radius
$R_\mathrm{core}$ (note that Equation (3) does not depend on this quantity.) If
the forming planet consists of a heavy-element core plus even a small
amount of gas bound in an envelope, the capture rate of planetesimals is
enhanced by the effects of gas drag and ablation as the accreting
objects pass through the envelope. We account for the enhancement of the
accretion rate by replacing $R_\mathrm{core}$ by $R_\mathrm{capt}$, the
effective capture radius. This quantity is determined at every time step
by integration of the orbits of planetesimals as they approach the 
planet at various impact parameters and pass through the envelope. The
details of how the procedure works are described in Pollack et al.~(1996)
based on earlier work by Podolak et al.~(1988).
The capture radius increases noticeably as the assumed planetesimal size
decreases, and even for 100 km planetesimals, the ratio  $R_\mathrm{capt}/
 R_\mathrm{core}$ can be a factor 10 if the envelope is sufficiently
massive. This enhancement effect is included in most of the calculations
reported here.

\subsection{Planetesimal Ejection}
Planetesimals are not only accreted by the growing planet but can also be 
ejected from the planet's feeding zone. The effect of planetesimal ejection 
is more profound at large radial distances and is taken into account in our 
model. Following Ida \& Lin (2004) the planetesimal ejection rate is calculated
according to
\begin{equation}
f_\mathrm{cap}=\frac{\text{accretion rate}}{\text{ejection rate}}= 
\bigg(\frac{2GM_{\odot}/a}{GM_\mathrm{core}/R_\mathrm{core}}\bigg)^2
\end{equation}
where $M_{\odot}$ and $M_{\text{core}}$ are the masses of the Sun and planetary core, 
respectively  (in this
case, $R_\mathrm{core}$ is the actual core radius, not an enhanced capture radius).
The probability that a planetesimal will be scattered,
rather than accreted, is then $1/(f_\mathrm{cap} + 1)$.  As Dodson-Robinson \& 
Bodenheimer (2010) show, this probability approaches unity when $M_\mathrm{core} \approx
10$ M$_\oplus$ and $a > 15$ AU. Thus it is an important factor in
limiting the final mass of the planet. In fact the use of Equation (6) gives
 a lower bound on the
actual scattering rate, as it takes into account only very close encounters and
scatterings directly to unbound states. Future work should take scattering into
account in a more detailed way.

\subsection{Available Formation Timescale}
The available time for forming a giant (or an {\it icy}) planet is typically taken 
to be the lifetime of the protoplanetary (gaseous) disk. At this point the gas disk
dissipates and the planetary growth is terminated, at least, in terms of the gaseous 
component. 
The lifetimes of protoplanetary disks are derived from observations and are of the 
order of several million years.  The median disk lifetime was found to be 3 Myr, 
a value we adopt in our model. However, some disks can survive on much longer 
timescales and can exist even up to 10 Myr (Hillenbrand 2008). In 
addition, the lifetime of gas in planetary disks can be longer than that of 
the dust, which can provide additional time for gas accretion. As a result, in 
some cases in which the planetary growth is limited by the disk's lifetime, we allow 
the longer formation timescale and investigate the sensitivity of the formation model 
to the assumed available formation timescale. 
As we present below in several cases, the timing on which the gas disappears has a 
major impact on the final mass of the planets, and it can determine whether 
the planet will become a gas giant or not. 

\subsection{Solid-to-Gas Ratio}
In addition to the final mass of the planets, which should be the same as the masses 
of Uranus and Neptune, formation models should also be able to lead to the correct 
solid-to-gas ratios within the planets. Although there is a fairly large uncertainty 
in the compositions of Uranus and Neptune (e.g., Fortney \& Nettelmann 2010), there are several constraints on the fraction of hydrogen and helium 
which they can contain. As discussed in Guillot (2005, and references therein), under the 
assumption that the envelopes of the planets consist of ices and rocks, upper limits 
for the hydrogen and helium masses in Uranus and Neptune can be derived, and those 
masses are found to be $\approx$ 4.2 M$_{\oplus}$ and $\approx$ 3.2 M$_{\oplus}$, respectively.  
A lower bound of $\approx$ 0.5 M$_{\oplus}$ for both Uranus and 
Neptune is inferred under the 
assumption that the outer envelope is pure hydrogen and helium. 

\section{Results}

The parameters assumed for the various runs are listed in Table 1.
Selected results are shown in Table 2, and the planet evolutions for
most cases are plotted in Figures 1--7. The starting times for the
plots in all cases are arbitrarily set at $\approx 2 \times 10^5$ yr, 
corresponding to the time required to build the initial core. However, 
the actual starting times will depend on $a$ and $\sigma_s$. For
example, Equation (8) of Rafikov (2011), which does not include the
effects of scattering or enhancement of the capture radius by the
presence of the envelope, would give  starting times of $3 \times
10^4$ yr for Run 12UN2 (low $a$, high $\sigma_s$)  to  $4 \times 10^6$
yr for Run 30UN1 (high $a$, low $\sigma_s$). These time estimates
are based on a solid accretion rate [dM/dt]$_{HIGH}$. 

\subsection{Formation at 20 AU}
We first model the formation of a planet at radial distance of 20 AU, which 
corresponds to {\it in situ} formation for Uranus or to 
Neptune's formation within the framework of the Nice model (Run 20UN1).
Based on the MMSN model, at 20 AU $\sigma_s  \approx $ 0.35 g cm$^{-2}$. 
The planetary core accretion rate is taken to be the maximum accretion rate 
[dM/dt]$_{HIGH}$ (Eq. 2), the enhanced planetesimal  capture radius is taken
into account with 1 km planetesimals,  and planetesimal scattering is included. 
The planetary growth under these assumptions is shown in Figure 1. The red, 
blue, and black curves represent the mass of the gaseous envelope (mostly 
hydrogen and helium), core (heavy elements), and total planetary mass, respectively. 

The upper panel shows the formation up to about 3 Myr. Under these conditions, 
the planetary growth is slow and the final planetary mass is 5 M$_{\oplus}$, 
significantly smaller than the mass of Uranus. If the gas disk dissipates at 
3 Myr, the forming planet will become a "mini-Neptune" with $M_\mathrm{core} = 3.6$
M$_\oplus$ and $M_\mathrm{env} = 1.4 $ M$_\oplus$ . However, if the disk lifetime 
is longer the planet can continue to grow. We then continue the formation to a 
longer timescale (see middle panel). As can be seen from the figure, at 8 Myr 
the planet can reach a mass exceeding 60 M$_{\oplus}$. Clearly, in this case 
the lifetime of the disk plays a crucial role in the formation process.  The 
calculation illustrates the point that even with $M_\mathrm{core} < 4$
M$_\oplus$ the planet can reach crossover mass ($M_\mathrm{core} = M_\mathrm{env}$), given enough time. 
 Crossover is actually obtained at 5.14 Myr at $M_\mathrm{core}=3.84 $ M$_\oplus$.

Finally, we search for the timescale for the formation of a planet with 
Uranus's mass  under these conditions. The result is shown in the 
bottom panel of the figure. To reach the mass of Uranus the gas disk has 
to dissipate at $\approx$ 7 Myr, and at $\approx$ 7.3 Myr in order to form a planet 
with Neptune's mass. It should be noted, however, that the final composition of the 
planets, assuming gas dissipation at about 7 Myr, is significantly different 
from that of Uranus/Neptune. The heavy element mass is only 5 M$_{\oplus}$,
while the rest is gas. Therefore, the forming planets will become something 
that is more like a "Mini-Saturn" rather than  a Uranus/Neptune-like object. \par

As discussed in Dodson-Robinson \& Bodenheimer (2010), the actual densities of 
the disk can be much higher than the ones derived from the MMSN model. In fact, 
the disk model of Dodson-Robinson et al.~(2009) suggests that in the outer region 
of the disk, the solid surface densities are higher by a factor of 10. We therefore 
repeat the calculation at 20 AU  (Run 20 UN2) but this time assume $\sigma_s=3.5$ g cm$^{-2}$ 
as derived by their disk model. Other parameters and assumptions are the same as
in the run with  $\sigma_s=0.35$ g cm$^{-2}$.  The results are shown in the top panel 
of Figure 2.  In that 
case the core accretion rate is higher which leads to a rapid growth. After 
only 0.55  Myr crossover mass is reached, at 16 M$_\oplus$,  and the planet can 
grow rapidly in mass. The simulation is stopped when the planet has reached 38 M$_{\oplus}$ 
but gas accretion is expected to continue. We can conclude that the combination 
of high core accretion rates with relatively solid-rich disks leads to the formation 
of giant gaseous planets within relatively short timescales. Note, however, 
that $M_\mathrm{core}$ levels out at about 16  M$_\oplus$. The isolation 
mass, which gives approximately the amount of solid material in the
feeding zone of the planet, is given by $M_\mathrm{iso} = 1.56 \times 10^{25}
(a^2 \sigma_s)^{3/2}$ g, where $a$ is given in AU and $\sigma_s$ in cgs units.
In this case  $M_\mathrm{iso} \approx 136$ M$_\oplus$, but scattering,
particularly at high core mass, removes a large fraction of that material.
In fact at the end of the simulation, practically all solid material
available to the planet has either been accreted or scattered. 

We next model (Run 20UN3) the formation at 20 AU with $\sigma_s=0.7$ g cm$^{-2}$,  which 
is about twice the MMSN value. The core accretion rate is set to its 
maximum value, and both the effect of planetesimal scattering and cross section 
enhancement due to the planetary envelope are {\it not} included.  
The results are shown in the middle panel of Figure 2.  In that 
case, a planet similar to Neptune can be formed within 1.3 Myr. At that point 
the core mass is $\sim$ 15 M$_{\oplus}$  with an envelope of 4 $M_{\oplus}$. If 
the simulation continues to longer times  more gas is accreted and the forming 
planet will become a gas giant planet. Under these conditions, in 
order to form a Neptune-like planet the gas has to dissipate relatively early. 
Note, however, that the planetary growth is expected to be slower and the
final heavy-element mass is expected to be smaller when 
planetesimal scattering is included.  
Finally, we model, in Run 20UN4,  the planetary growth at 20 AU, $\sigma_s$ = 1.7 g cm$^{-2}$ but this time with the transitional  core 
accretion rate $[dM/dt]_{LOW}$ (Eq.~5). The results are presented in the bottom panel of Figure 2. 
Even with the low accretion rate it is found that a planet of similar 
mass to that of Uranus can be formed within less than 2 Myr; however, the gaseous mass is significantly higher than that of Uranus. 
\par

Clearly, the properties of the planet can change significantly even due to changes 
in the solid-surface density alone. We can also conclude that the high accretion 
rates can now allow for {\it in situ} formation at 20 AU, even when the MMSN solid 
surface densities are used. However in that case, with very low surface density, the
conclusion is marginal. First, as mentioned, the time required to reach Uranus
mass is 7 Myr, which is beyond the lifetime of most gas disks. Second, to allow
the use of the maximum accretion rate, the embryo has to grow big enough to excite
the planetesimal velocities so that they collide, fragment, and produce the
required small particles.  Based on the work of Rafikov (2003, 2004),	 at 20 AU and in the MMSN, 
 the required
embryo mass is of order $10^{24}$ g, and the time to build that embryo is very
roughly estimated at $10^6$ to $10^7$ yr. Therefore a detailed numerical simulation
of the planetesimal dynamics, including fragmentation,  is required to determine whether the use of the
high accretion rate is justified in this particular case. At smaller distances from the Sun,  or
with higher values of $\sigma_s$, the
corresponding time is considerably shorter. 

\subsection{Formation at 15 AU}
In this section we investigate the formation of a planet at 15 AU. 
First (Run 15UN1), we consider $\sigma_s$ = 0.55 g cm$^{-2}$, close to the value in the MMSN. The core accretion rate is set to [dM/dt]$_{HIGH}$, scattering of planetesimals
is included, and envelope enhancement is included. 
The results are shown in the top panel of Figure 3.  With this low surface density 
the planetary growth is slow and at $\sim$ 2.5 Myr the planet has a core mass of 
3.6 M$_{\oplus}$ and an envelope mass of 0.8 M$_{\oplus}$.  At  9.7 Myr crossover 
mass is still not reached, and the planet has a core mass of about 4 M$_{\oplus}$ 
and an envelope of  3.1 M$_{\oplus}$. Clearly, under these assumptions it is not 
possible to form a Neptune/Uranus mass planet. It should be noted, however, that 
at the point the simulation stops, the gas accretion rate is about 
0.3 M$_{\oplus}$/Myr, 
therefore a slightly higher solid surface density and/or longer disk lifetime 
would probably lead to the formation of a gaseous planet. \par 
   
We next model the planetary formation at 15 AU with $\sigma_s$ = 5.5 cm$^{-2}$
(Run 15UN2). 
Other assumptions are the same as in the previous case.
It is found that the crossover mass is reached already after $\approx$ 6$\times$10$^5$ 
years when the core and envelope masses are 18.9 M$_{\oplus}$. The results 
for this case are shown in the bottom panel of Figure 3. Clearly, this combination of parameters is more preferable for giant planet formation than formation of intermediate-mass planets. 

The use of  [dM/dt]$_{HIGH}$ implies that small planetesimals, less than 100 meters in size,
dominate the accretion (Rafikov 2004). Our calculations with 1 km planetesimals are not 
entirely consistent; therefore we run a calculation with planetesimal size 50 meters
and otherwise with the parameters of Run 15UN1.  In our calculation the only effect of
decreasing the planetesimal size is to increase the enhancement of the cross section
for planetesimal capture. Thus the the phase during which core accretion dominates is
shortened; this phase is short in any case. 
Once the core mass approaches its limiting value, and envelope accretion is
the dominant process, the effect of the smaller planetesimals is very small. 
In Run 15UN3, the limiting core mass is approached only 2.2 $\times 10^5$ years after the
start, while in Run 15UN1 the corresponding time is 3.25 $\times 10^5$ years. 
In Run 15UN3 at 2.5  Myr, the core mass is 3.62 M$_\oplus$ and the envelope mass is  0.85 M$_\oplus$, 
practically the same as in Run 15UN1.     The calculation in 15UN3 is run to
9.7  Myr, as in Run 15UN1, and at that time the core mass is 4.03   M$_\oplus$ and
the envelope mass is 3.65  M$_\oplus$. These results give the same conclusion as in 
Run 15UN1: with these assumptions it is not possible to produce a Uranus or Neptune mass
planet in a reasonable time.

\subsection{Formation at 12 AU}
We next model the formation of a planet at 12 AU, the minimum radial 
distance for the formation of Uranus/Neptune as suggested by the Nice model. 
At 12 AU the solid surface density is relatively high, providing preferable 
conditions for planet formation.
 
First (Run 12UN1), we consider a solid surface density of 
$\sigma_s$ = 0.75 g cm$^{-2}$,  which is 
approximately that of the MMSN at that radial distance,   and assume  the 
high solid accretion rate (Eq.~2). As in previous cases, the enhancement of the capture 
rate is included, with 1 km planetesimals. The results are 
shown in the top panel of Figure 4. 
The planetary growth is somewhat similar to that  derived for the low-density case 
at 20 AU. The core mass reaches $\approx$ 3.5 M$_{\oplus}$ relatively fast and then 
levels off, while the envelope mass reaches 0.65 M$_{\oplus}$ at 2.5 Myr. 
Clearly, under these conditions it is not possible to form a planet which resembles 
Uranus/Neptune. In order to demonstrate whether it is possible  to form a 
Uranus/Neptune-like planet we continue the simulation up to 10 Myr. At that time
we obtain $M_\mathrm{core} = 3.8 $ M$_\oplus$ and $M_\mathrm{env} = 2.2 $ M$_\oplus$. 
We can therefore conclude that under such conditions it is not possible to form a 
planet of the order of 15 M$_{\oplus}$; instead a Super-Earth or Mini-Neptune 
planet is formed. Comparing Runs 20UN1 and 12UN1, both runs having
low surface densities, the former case develops a somewhat higher core
mass, because more solid material is available in the disk at the
larger distance. The envelope accretion rate is strongly dependent on
the core mass (Pollack et al. 1996), so 20UN1 is able to reach
crossover in  $< 10$ Myr, while 12UN1 is not.  \par

We next consider planet formation 
 with $\sigma_s$ = 7.5 g cm$^{-2}$  which corresponds to $\approx$ 10 times MMSN
(Run 12UN2). 
Other assumptions are the same as in the previous case.
The results are shown in the lower  panel of Figure 4. Under these conditions the growth of 
the planet is rapid,  and 
 at  0.52 Myr it reaches  the crossover mass  of 21  M$_{\oplus}$
 which leads to rapid gas accretion. The simulation was terminated at 0.56 Myr with a
total mass of  48 M$_\oplus$ and an envelope mass of $\approx$ 26 M$_{\oplus}$. 
If the disk lifetime under these conditions is of the order of several Myr, 
a gaseous planet is expected to form. This simulation is similar to the one at 
15 AU with the high solid-surface density. In both cases, a giant planet can be formed 
quickly. By the end of both runs most of the planetesimals in the feeding zone 
have either been accreted onto the core or scattered. However, closer to 
the Sun, 
a larger fraction of the planetesimals can be accreted; while at 12 AU $f_{cap}$ = 0.18 
at the end of the simulation, at 15 AU, $f_{cap}$ = 0.13.  
Thus  $M_\mathrm{core}$ at
crossover is somewhat smaller at 15 AU (18.9 vs. 21 M$_\oplus$), and because of 
the lower surface density at 15 AU, the time to reach crossover is somewhat longer
($6.2 \times 10^5$ vs.  5.2 $\times 10^5$ yr.)\par

We next consider (Run  12UN3) the high solid surface density case (i.e., $\sigma_s$ = 7.5 g cm$^{-2}$) 
but this time using the ``transition"  accretion rate [dM/dt]$_{LOW}$ (Eq.~5).  
Other parameters are the same as in the previous case.
The results are shown in the top panel of Figure 5. Even under these conditions 
the planetary growth is fairly rapid and after about 0.9 Myr the planet has 
reached a total mass of $\approx$ 45 M$_{\oplus}$ with an envelope mass of
24 M$_\oplus$. This case is very similar 
to the previous case using the maximum accretion rate,  with the main difference 
being the timescale to reach the crossover mass. Nevertheless, due to the 
high solid surface density, a giant planet can be formed within the lifetime 
of most protoplanetary disks. In these cases, the challenge is to form 
planets with final masses similar to those of Uranus and Neptune. 
If Uranus/Neptune indeed formed at 12 and/or 15 AU,  they can provide important 
constraints on $\sigma_s$ and the accretion rates at these locations, since 
according to our simulations, the growing planets can become gaseous planets fairly easily.  \par
This case was recalculated with a planetesimal size of 100 km rather than 1 km,
again using [dM/dt]$_{LOW}$ (not plotted or listed). The end result was practically the same, namely,
the onset of rapid gas accretion, with a crossover mass of about 21 M$_\oplus$.
The time to reach crossover, however was twice as long, 1.8 Myr instead of 0.9 Myr,
as a result of the reduced enhancement of the planetesimal capture cross section
with the larger planetesimals.

The case discussed in the previous paragraph is somewhat similar to a calculation done
by Dodson-Robinson \& Bodenheimer (2010), who considered planet formation at 12 AU
with $\sigma_s=8.4$ g cm$^{-2}$. In their calculation, planetesimal scattering was
included but the core accretion rate was a factor 2 lower than [dM/dt]$_{LOW}$. Their
result shows that Uranus mass is reached at 4 Myr with $M_\mathrm{core} = 13.5$ M$_{\oplus}$
and $M_\mathrm{env} = 1.0$  M$_{\oplus}$. Neptune mass is reached at 4.2 Myr with
$M_\mathrm{core} = 15$  M$_{\oplus}$ and $M_\mathrm{env} = 2$  M$_{\oplus}$. However, if
the disk does not dissipate until later times, accretion can continue up to much 
higher masses. We have redone this simulation with the same parameters but with an
improved version of the formation code.  We take a planetesimal size of 100 km
to be consistent with their assumptions. 
Our results show that Uranus mass is reached
at 1.76 Myr with  $M_\mathrm{core} = 13.9$  M$_{\oplus}$ and $M_\mathrm{env} = 0.6$  M$_{\oplus}$.
Neptune mass is reached at 1.89 Myr with $M_\mathrm{core} = 16$  M$_{\oplus}$
and $M_\mathrm{env} = 1$  M$_{\oplus}$. Again, accretion continues to higher masses.
Our times are about a factor 2 shorter than those of  Dodson-Robinson \& Bodenheimer (2010), 
mainly because the envelope enhancement of the planetesimal capture cross section is 
more accurately calculated in our case and turns out to be higher.
\par

Finally (Run 12UN4), we model the formation of a planet at 12 AU, 
with the transitional accretion rate but with lower $\sigma_s$ than in 
Run 12UN3. 
The results are shown in the bottom panel of Figure 5. Although the 
formation timescale is longer compared to that in Run 12UN3, it is found that  
crossover mass can be reached within $\approx$ 2.5 Myr. The value of the
crossover mass is, however, almost a factor 2 lower than in Run 12UN3.
At this point the planetary mass is over 20 M$_{\oplus}$,  and 
the solid-to-gas ratio is close to  1.

\subsection{The Effect of Planetesimal Scattering - A Test At 30 AU}

The farthest formation location so far considered in this work is 20 AU, 
which is typically considered as an upper bound for Uranus and Neptune 
based on the Nice model. Formation at larger radial distances is however possible. 
Since the surface density decreases with radial distance the formation process is 
somewhat less efficient and the forming planets are expected to have smaller 
masses. Nevertheless, with high core accretion rates, and with sufficiently
high $\sigma_s$ even at 30 AU planets 
could be formed, and even become gas giants. Our simulations, however, are
limited to the case where $\sigma_s$ is close to the value for the MMSN. 

It should be noted, however, that our simulations start with solid cores of 
$\approx$ 1.3 M$_{\oplus}$. At 30 AU, the timescale for forming such cores 
can be a few million years, which delays the formation process significantly. 
In addition, planetesimal scattering is more important for large 
radial distances. In this section we model planetary formation at 30 AU. 
In order to demonstrate the importance of planetesimal scattering 
and its crucial effect on the planetary growth, we compare two cases of 
planet formation at 30 AU (Runs 30UN1 and 30UN2),  which differ 
solely by this assumption. The solid surface density 
is assumed to be  low, 0.2 g cm$^{-2}$, and  the high accretion rate 
[dM/dt]$_{HIGH}$ is used.  The results are shown in Figure 6. The top 
panel presents the planetary growth when planetesimal scattering is 
included. It is found that after 3 Myr, a planet with a total mass of 
$\approx$ 5 M$_{\oplus}$ is formed. The gaseous mass by the end of the 
simulation is 1.6 M$_{\oplus}$. The planetary growth is somewhat 
similar to that in Run 20UN1,   at 20 AU with the low $\sigma_s$ (Figure 1). 
If the planet is allowed to grow on a longer timescale,  crossover mass is 
expected to be reached before 10 Myr,  leading to the formation of a giant 
or intermediate mass planet within 10 Myr. The total mass of heavy elements 
is, however, small, and therefore,  regardless of the final mass of the planet,  it is expected to be gas-dominated.  The results for the planetary growth 
when planetesimal scattering is {\it not} considered (Run 30UN2) are shown 
in the bottom panel. In this case the planet can grow faster, and crossover 
is reached after $\approx$ 2.5 Myr with a higher solid mass than in 
Run 30UN1. In that case Uranus/Neptune-mass planets can be formed even at 30 
AU. For the case without planetesimal scattering,  even if we add the 
$\approx$ 4.4 Myr that are required to build the seed core, 
intermediate-mass and gas giant planets can be formed at 30 AU 
when disk lifetimes are of the order of 10 Myr. The solid-to-gas ratio 
by the end of the simulation is close to one. Clearly, planetesimal 
scattering plays a major role in planet formation models, when relatively 
large radial distances are considered, and it has a crucial effect on both 
the formation timescale and the final composition of the planets. 

However, at a distance of 12 AU the effect is much less important. A
comparison to Run 12UN1 was run without scattering (not plotted or listed). 
The run  without scattering also did not reach crossover; at 7 Myr
it reached $M_\mathrm{core} = 5.05$ M$_\oplus$ and $M_\mathrm{env} =
2.84$ M$_\oplus$, in comparison with  $M_\mathrm{core} = 3.7$ M$_\oplus$ 
and  $M_\mathrm{env} = 1.62 $  M$_\oplus$ at the same time in
Run 12UN1.

\subsection{The Effect of the Planetesimal Size}

Clearly, since small planetesimals are more affected by the gaseous envelope,
the planetary growth is more efficient when small planetesimals are assumed. 
We next present two cases (Runs 20UN5 and 12UN5) in which the planetesimals 
are assumed to be relatively large, i.e. 100 km (see also Section 3.2, 
Run 15UN3).  Results for the two 
runs with 100 km-sized planetesimals are shown in Figure 7. The top panel 
(Run 20UN5) corresponds to formation at 20 AU with the maximum accretion 
rate and $\sigma_s=0.35$ g cm$^{-2}$, i.e., similar to the case presented 
in Figure 1 but assuming 100 km-sized planetesimals instead of 1 km. 
By comparing the two cases it can be concluded that although the growth 
is slightly slower in Run 20UN5, the overall growth of planet after 3 Myr 
is very similar. At 3 Myr the core mass is found to be 3.57 M$_{\oplus}$ 
and the envelope mass  1.6 M$_{\oplus}$, compared to $M_\mathrm{core}
= 3.55$ M$_\oplus$ and $M_\mathrm{env} =  1.43$ M$_\oplus$ in 
Run 20UN1 at the same time. The main difference between these two 
cases is that when larger planetesimals  are used, it takes a somewhat
longer time to build up the core during the earlier phases to the (scattering limited) mass of $\sim$ 3.5 M$_{\oplus}$. \par

The bottom panel of Figure 7 illustrates  Run 12UN5, at 12 AU with
$\sigma_s=3$ g cm$^{-2}$, [dM/dt]$_{LOW}$, and 100 km planetesimals. This
case should be compared with Run 12UN4 (bottom panel of Figure 5), which
has the same parameters except for a planetesimal size of 1 km; crossover
mass of 11.6 M$_\oplus$ is reached after 2.3 Myr. With 100 km planetesimals,
a Uranus-mass planet can be formed within about 3.3 Myr, with core mass
10.5  M$_\oplus$ and envelope mass 4.0  M$_\oplus$. After that point
the gas accretion rate is significantly higher than the core accretion
rate; as a result a planet with Neptune's mass can be formed under
these conditions after 3.6 Myr, with $M_\mathrm{core} = 10.9 $ M$_\oplus$ 
and $M_\mathrm{env} =  6.1 $ M$_\oplus$. The solid-to-gas ratio is too
low compared with that of Neptune; however the case with the higher
planetesimal size leads to a higher solid-to gas ratio at times around
3 Myr, since crossover mass is not reached by that time.

\subsection{The Lambrechts--Johansen (LJ) Accretion Rate}
The LJ rate (Eq. 3) is higher than [dM/dt]$_{HIGH}$ by roughly
one order of magnitude, depending on the stage of evolution. 
The cases with high $\sigma_s$, 
which include Runs 20UN2, 15UN2, 12UN2, and 12UN3, all reach crossover
mass in relatively short times. If these runs were to be redone with
the LJ rate, they would simply reach crossover mass faster, with 
similar core masses; the core masses are determined by the amount
of solid material available and by the effects of scattering, not
by the accretion rate. On the other hand, the low $\sigma_s$ cases, 
which include Runs 20UN1, 15UN1, 12UN1, 30UN1, and 20UN5, do not
reach crossover in 3 Myr, although they could over longer times.
The core masses, which fall in the range 3--4 M$_\oplus$ at 3 Myr,
are limited by the isolation mass, which increases as $a^{3/4}$ in the
MMSN, 
 as reduced by the effects of scattering, which become more important
at larger $a$. The gas accretion
rates at these low core masses are slow, and are little affected by the
core accretion rate, once the limiting core masses have been reached.	
Thus the use of a faster core accretion rate would simply speed up
the time to approach the limiting core mass but would have little
effect on the results at longer times.

 A test calculation has been
run at 20 AU and $\sigma_s = 0.35$ g cm$^{-2}$, with the LJ rate.
 A core mass of 3.2 M$_\oplus$ is reached after $5 \times 10^4$ years
after the start, as compared with about $3 \times 10^5$ yr in the
comparison run 20UN1 (top panel of Figure 1). The end result for
the test case after 3 Myr is $M_\mathrm{core}= 3.65$ M$_\oplus$
and $M_\mathrm{env}= 2.13$ M$_\oplus$, while for Run 20UN1 the
corresponding numbers are 3.59 and 1.72 M$_\oplus$, respectively. 
Thus the basic conclusions of this paper are only weakly affected
by the choice of Equation (3) rather than Equation (2).

A problem may arise, however, in the use of the LJ theory.
The time to reach the transition mass, Equation (4), can be quite
long. In fact the LJ estimate of the time to reach this mass
(their Eq. 42) exceeds 10$^8$ yr at 20 AU. Thus fast accretion
by this mechanism requires, first, that a significant fraction
of the solid material be in the form of ``pebbles" (cm-size particles)
close to the midplane, and second, that at least some planetesimals close to
the transition mass ($\approx$ 1000 km  in size) must have formed
early by some independent process (Lambrechts \& Johansen 2012).

\section{Conclusions}

Understanding the formation of Uranus and Neptune is crucial for understanding 
the origin of our solar system, and in addition, the formation of intermediate-mass planets around other stars.  
Planets that are similar to Uranus and Neptune in terms of mass are likely to form 
by core accretion, i.e., from a growing solid core which accretes gas at a lower rate, 
although alternative mechanisms should not be excluded 
(Boss et al. 2002; Nayakshin 2011). 
If indeed formed by core accretion, such planets must form fast enough to ensure 
that gas is accreted onto the core, but  at the same time slow enough, in order to remain small in 
mass and not become gas giant planets. \par

Our study shows that simulating the formation of Uranus and Neptune is not 
trivial, and that getting the correct masses and solid-to-gas ratio depends 
on the many (unknown) model parameters. 
Even small changes in the assumed parameters can lead to a very different 
planet. The core accretion rate and the disk's properties such as solid-surface density and 
the planetesimals' properties (sizes, dynamics, etc.) play a major role in the formation process, and even 
small changes in these parameters can influence the final masses and 
compositions of the planets considerably. \par

We have used high accretion rates for the solids and have investigated their 
impact on the planet formation process. With these high accretion rates  
and with values of $\sigma_s$ about 10 times those in the MMSN, 
the 
formation timescale problem for formation of Uranus or Neptune at 20 AU disappears. 
However, a new problem arises - the formation of the planets can be so efficient 
that instead of becoming failed giant planets, Uranus and Neptune would 
become giant planets, similar to Jupiter and Saturn.
The situation becomes even worse with formation at smaller radial distance 
where the solid surface density is high. At radial distances such as 12 and 15 AU, 
the planets reach runaway gas accretion within a timescale which is shorter 
than the average lifetimes of protoplanetary disks, leading to the 
formation of gaseous giant planets.  

However, if the values of $\sigma_s$
appropriate for the MMSN are used, a different problem arises. Even with
high core accretion rates, the resulting solid masses are too low 
compared with those of Uranus/Neptune, {at all distances.  Scattering of
planetesimals is an important effect in this regard.} Our results suggest that
intermediate values of $\sigma_s$, perhaps combined with relatively
low core accretion rates (e.g.~Run 12UN5), are needed to satisfy the
joint constraints provided by disk lifetimes, total masses, and 
solid-to-gas ratios. 
\par
 
It should be noted that the ``true" core accretion rate  to form Uranus 
and Neptune is not known and could be different from the values we consider 
here. Our work shows that even with a relatively low accretion rate,  combined with a  relatively high
solid surface density $\sigma_s$, it is possible to form a giant
planet in a relatively short time (Run 12UN3). However, different, but still reasonable
choices for these parameters can produce quite different results; for example Run 12UN1
produces a planet of less than half the mass of Uranus after 10 Myr. Finding 
the right set of parameters that will lead to the formation of planets 
with the final masses of Uranus and Neptune is challenging, and getting the 
correct gas masses is even harder. In some cases the gas accretion rate becomes 
high enough and an increase in mass  to the Uranus/Neptune mass range 
occurs, but the solid-to-gas ratio 
in the models  is inconsistent with that of Uranus and Neptune (see Table 2). 
The cases we present simply demonstrate the sensitivity of the planetary 
formation to the assumed parameters, and while they do provide possible 
scenarios for the formation of Uranus and Neptune, they are certainty not 
unique. Clearly, the core accretion rate is a major uncertainty in 
planet formation models,  and a better determination of this property 
(and its time evolution) will have a significant impact on simulations 
of planetary growth of both terrestrial and giant planets.
\par

Our work suggests that under the right conditions,  
{\it in situ} formation for Uranus at 20 AU, or formation of Neptune at about
the same distance,  is possible. At 30 AU,  our run with $\sigma_s$ near the
value for the MMSN and with planetesimal scattering, showed that a Neptune-mass
planet was not formed. Although Rafikov (2011) shows that the use of the high
accretion rate should result in evolution to rapid gas accretion out to 
40--50 AU in a MMSN in 3 Myr, our result is different. The main reason is 
that planetesimal scattering limits the core mass to about 3.3 M$_\oplus$, leading
to a slow accretion rate for the envelope. However that result could change if a higher value 
of  $\sigma_s$ were taken.  

{\it In situ} formation at relatively large radial distances seems to be possible 
for intermediate-mass 
planets in general. In addition, high solid 
surface density, which is expected in metal-rich environments, can 
lead to fast core formation, and therefore to the formation of giant planets 
instead of intermediate-mass planets. The latter provides a natural 
explanation to the correlation between stellar metallicity and the 
occurrence rate of gas giant planets. Nevertheless, we also find that giant planets can be formed in low-metallicity environments.  As a result, giant planets around low-metallicity stars 
should not necessarily be associated with formation by gravitational instability.  
It is also concluded that the formation of Uranus/Neptune-mass planets is not always challenging in terms of the formation 
timescale but often, in terms of reducing the gas accretion in order to prevent the formation of gaseous planets. 
\par

Finally, while our work emphasizes once more the difficulty to simulate 
the formation of the solar-system planets very accurately, it 
provides a natural 
explanation for the diversity in planetary parameters in extrasolar planetary systems. 
Since planetary disks are expected to have different physical properties (e.g., surface densities, 
lifetimes) it is clear that the forming planets will have different 
growth histories, as well as different final masses and compositions.

\subsection*{Acknowledgments} 
P. B. was supported in part by a grant from the NASA program ``Origins
of Solar Systems".

\newpage
\section*{REFERENCES}
\begin{enumerate} 
\item[]{} Boss, A. P., Wetherill, G. W.~\& Haghighipour, N. 2002.  
Icarus, 156, 291 
\item[]{} D'Angelo,  G., Durisen,  R. H.~\& Lissauer,  J. J. 2011. In: 
Exoplanets, ed. S. Seager (Tucson: Univ. of Arizona Press),  319 
\item[]{} Dodson-Robinson, S. E.~\& Bodenheimer, P. 2010.  Icarus, 207, 491
\item[]{} Dodson-Robinson, S. E., Willacy, K., Bodenheimer, P., 
Turner, N. J.,  \& Beichman, C. A. 2009.  Icarus, 200, 672
\item[]{} Fortney,  J. J.~\& Nettelmann, N. 2010. Space Sci. Rev., 
 152, 423 
\item[]{} Goldreich, P., Lithwick, Y.~\&  Sari, R. 2004.  
Annu. Rev. Astron. Astrophys., 42, 549
\item[]{} Guillot, T. 2005. Annu. Rev. Earth  Planet. Sci., 33, 493
\item[]{} Hahn, J. M.~\&  Malhotra, R. 1999. \aj, 117, 3041
\item[]{} Helled,  R., Anderson,  J. D., Podolak,  M.~\& Schubert, G. 
 2011. \apj,  726, 15
\item[]{} Hillenbrand,  L. 2008. Phys. Script., 130, 014024. 
\item[]{} Ida, S.~\& Lin,  D. N. C. 2004. \apj, 604, 388
\item[]{} Kley, W.~\& Nelson, R. 2012.  Annu. Rev. Astron. Astrophys., 
50, 211
\item[]{} Lambrechts, M.~\& Johansen, A. 2012. A\&A, 544, A32
\item[]{} Levison, H. F., Duncan, M. J.~\&  Thommes, E. W. 2010, \aj, 139, 1297
\item[]{} Levison, H. F.~\& Morbidelli, A. 2007, Icarus, 189, 196 
\item[]{} Lissauer,  J. J., Hubickyj, O., D'Angelo, G.~\& Bodenheimer, P. 
 2009,  Icarus, 199, 338
\item[]{} Malhotra, R. 1995, \aj, 110, 420
\item[]{} Movshovitz, N., Bodenheimer, P., Podolak, M.~\& Lissauer, J. J.
2010. Icarus, 209, 616
\item[]{} Nayakshin, S. 2011. MNRAS, 416, 2974
\item[]{} Nettelmann, N., Helled, R., Fortney, J. J.~\&  Redmer, R. 2013.  
Planet. Space Sci., 77, 143 
\item[]{} Podolak, M., Pollack, J. B.~\& Reynolds, R. T. 1988.  Icarus,
73, 163
\item[]{} Pollack, J. B., Hollenbach, D., Beckwith, S. et al. 1994.  
\apj, 421, 613 
\item[]{} Pollack, J. B., Hubickyj, O., Bodenheimer, P. et al.~1996.  
Icarus, 124, 62. 
\item[]{} Rafikov,  R. R.  2003. \aj,  125, 942 
\item[]{} Rafikov,  R. R.  2004. \aj,  128, 1348 
\item[]{} Rafikov,  R. R.  2011. \apj,  727, 86
\item[]{} Rogers, L. A., Bodenheimer, P.,  Lissauer, J. J.~\& Seager, S.~2011. \apj, 738, 59 
\item[]{} Safronov,  V. S. 1969. Evolution of the Protoplanetary Cloud 
and Formation of the Earth and Planets (Moscow: Nauka), in Russian. English 
translation: NASA--TTF--677, (Jerusalem: Israel Sci. Transl. 1972)
\item[]{} Thommes, E. W., Duncan, M. J.~\& Levison, H. F. 1999. Nature, 
402, 635  
\item[]{} Tsiganis, K.,  Gomes, R., Morbidelli A.~\& Levison, 
H. F.~2005. Nature, 435, 459
\item[]{} Weidenschilling,  S. J. 1977.  Astrophys. Space Sci., 51, 153 
\end{enumerate} 


\begin{sidewaystable}[h!]
\begin{center}
{\renewcommand{\arraystretch}{1.}
\vskip 11pt
\begin{tabular}{lc c c c c c c c|}
\hline
\\
RunName & Radial Distance (AU) & $\sigma_s$ (g cm$^{-2}$) & P-Size(km) 
& P-Scattering & $T_{\text {neb}} (K)$ & $\rho_{\text {neb}}$ (g cm$^{-3}$)
& $\dot M_\mathrm{core}$  
\\
\hline
20UN1 &20 & 0.35 & 1 & YES & 20 & $8 \times 10^{-13}$ & eq.~(2)\\
20UN2 & 20 & 3.5 & 1 & YES & 20 & $8 \times 10^{-12}$ & eq.~(2)\\
20UN3 & 20& 0.7  & 1 & NO  & 20 & $1.6 \times 10^{-12}$ & eq.~(2)\\
20UN4 & 20 & 1.7 & 1 & YES & 20 & $4 \times 10^{-12}$ & eq.~(5)\\
15UN1 & 15 & 0.55  & 1& YES  & 40 & $1.7 \times 10^{-12}$ & eq.~(2)\\
15UN2 &  15 & 5.5 & 1 & YES   & 40 & $1.7 \times 10^{-11}$ & eq.~(2)\\
15UN3 & 15 & 0.55 & 0.05 & YES   & 40 & $1.7 \times 10^{-11}$ & eq.~(2)\\
12UN1 & 12 & 0.75 & 1 & YES  & 50 & $3 \times 10^{-12}$ & eq.~(2)\\
12UN2 & 12 & 7.5  & 1 & YES  & 50 & $3 \times 10^{-11}$ & eq.~(2)\\
12UN3 & 12 & 7.5  & 1 & YES & 35 & $3 \times 10^{-11}$ & eq.~(5)\\
12UN4 & 12 & 3  & 1 & YES& 50 & $1.2 \times 10^{-11}$ &   eq.~(5)\\
30UN1 & 30 & 0.2 & 1& YES & 20 & $3 \times 10^{-13}$ &  eq.~(2)\\
30UN2 & 30 & 0.2  & 1 & NO & 20 & $3 \times 10^{-13}$ &  eq.~(2)\\
20UN5  & 20 & 0.35  & 100 & YES & 20 & $8 \times 10^{-13}$ &  eq.~(2)\\
12UN5  & 12 & 3  & 100 & YES & 50 & $1.2 \times 10^{-11}$ &  eq.~(5)\\
\hline
\end{tabular} 
}
\caption{\label{data1} 
The various cases considered in this work. $\sigma_s$ is the 
solid surface density at a given radial distance, ``P-Size" is the
planetesimal size, ``P-Scattering" indicates whether planetesimal 
scattering is included in the simulation,  $T_{\text {neb}}$ and 
$\rho_{\text {neb}}$ are the temperature and density of the disk at 
the formation location of the planet and $\dot M_\mathrm{core}$ is the assumed 
core accretion rate. 
}
\end{center} 
\end{sidewaystable}

\clearpage

\begin{sidewaystable}[h!]
\begin{center}
{\renewcommand{\arraystretch}{1.}
\vskip 11pt
\begin{tabular}{lc c c c c c c c c c|}
\hline
\hline
\\
RunName & $t_\mathrm{crossover}$ & $M_\mathrm{crossover} $ & $t_\mathrm{end}$
 & $M_\mathrm{core,end}$ & $M_\mathrm{env,end}$ & $M_\mathrm{tot,end}$ & $f_{solid-to-gas}$ &
 $\dot M_\mathrm{core,end}$ & $\dot M_\mathrm{env,end}$ \\
& Myr& M$_\oplus$ & Myr & M$_\oplus$ & M$_\oplus$ & M$_\oplus$ &  &
M$_\oplus$ yr$^{-1}$ & M$_\oplus$ yr$^{-1}$\\
\\
\hline
\hline
20UN1   & 5.14 & 3.83 & 7.9 & 5.6 & 60 & 65.6 & 0.09  & $1.5 \times 10^{-5}$ & 
$1.9 \times 10^{-4}$\\
20UN2  & 0.54   & 15.8 & 0.59 & 16.2  & 20.3 & 36.5 & 0.8  & $3.0 \times 10^{-5}$ &
$6.0 \times 10^{-5}$\\
20UN3 & $-$ & $-$ & 1.3& 14.3 & 3.9 & 18.2 & 3.67 & $3.3 \times 10^{-6}$&
$8.8 \times 10^{-6}$ \\
20UN4 & $-$ & $-$& 1.67 & 9.8 & 4.8 & 14.6 &  2.04 &$2.0 \times 10^{-6}$ &
$1.6 \times 10^{-5}$\\
15UN1 & $-$   & $-$ & 9.7 & 4.0& 3.1 & 7.1& 1.29  & $3.2 \times 10^{-8}$  &
$3.0 \times 10^{-7}$\\
15UN2 & 0.62  & 18.9 &0.63 & 19.4 & 26.2 & 45.6 &  0.74 & $3.3 \times 10^{-5}$ &
$1.3 \times 10^{-4}$\\
15UN3 & - & - & 9.7 & 4.03 & 3.65 & 7.7 &  1.10 & $3.2 \times 10^{-8}$ &
$4.0 \times 10^{-7}$\\
12UN1 & $-$  & $-$ &10.0 & 3.8& 2.2 & 6.0 & 1.73 & $2.3 \times 10^{-8}$ &
$1 \times 10^{-7}$ \\
12UN2 & 0.52  & 21.2 &0.56 &21.5 & 26.1 & 47.6 & 0.82 &  $8.1 \times 10^{-5}$ &
$4.2 \times 10^{-4}$\\
12UN3 & 0.89  & 20.8 & 0.90& 21.0 & 23.9 & 44.9& 0.88 &$5.5 \times 10^{-5}$ &
$4.1 \times 10^{-4}$ \\
12UN4 & 2.39  & 11.6 & 2.40 & 11.6& 11.7 & 23.3& 0.99 &$1.3 \times 10^{-6}$ &
$5.4 \times 10^{-6}$ \\
30UN1 & $-$ & $-$ & 3.1 & 3.3 & 1.6 & 4.9& 2.06 &$1.9 \times 10^{-7}$ &
$1.6 \times 10^{-5}$\\
30UN2 & 2.6  & 9.45 & 2.6 & 9.4 & 10.1& 19.5 & 0.93 & $1.5 \times 10^{-7}$ &
$4.7 \times 10^{-5}$\\
20UN5 & $-$  & $-$ & 2.9 & 3.6 & 1.6 & 5.2 & 2.25 & $2.7 \times 10^{-7}$ & 
$1.5 \times 10^{-6}$  \\
12UN5 & $-$ & $-$& 3.7 & 11.0 & 6.8 & 17.8 & 1.62 &$1.3 \times 10^{-6}$ &
$9.4 \times 10^{-6}$ \\
\hline
\end{tabular} 
}
\caption{\label{data2} 
Results:  
$t_\mathrm{crossover}$ is the time (if applicable) at which 
$M_\mathrm{core} = M_\mathrm{env}$, 
 $M_\mathrm{crossover}$ is $M_\mathrm{core}$ at that time, 
 $t_\mathrm{end}$ is the time at which the simulation stops,  
$M_\mathrm{core,end}$, $M_\mathrm{env,end}$ and $M_\mathrm{tot,end}$ 
correspond to the core, envelope, and total mass at the end of the run, 
respectively. $f_{solid-to-gas}$ is the solid-to-gas ratio, $\dot M_\mathrm{core,end}$ and $\dot M_\mathrm{env,end}$ 
are the core accretion rate and the gas accretion rate at the end of the simulation. 
}
\end{center} 
\end{sidewaystable}

\clearpage
\begin{figure}
\begin{center}
\includegraphics[angle=0,height=6.cm]{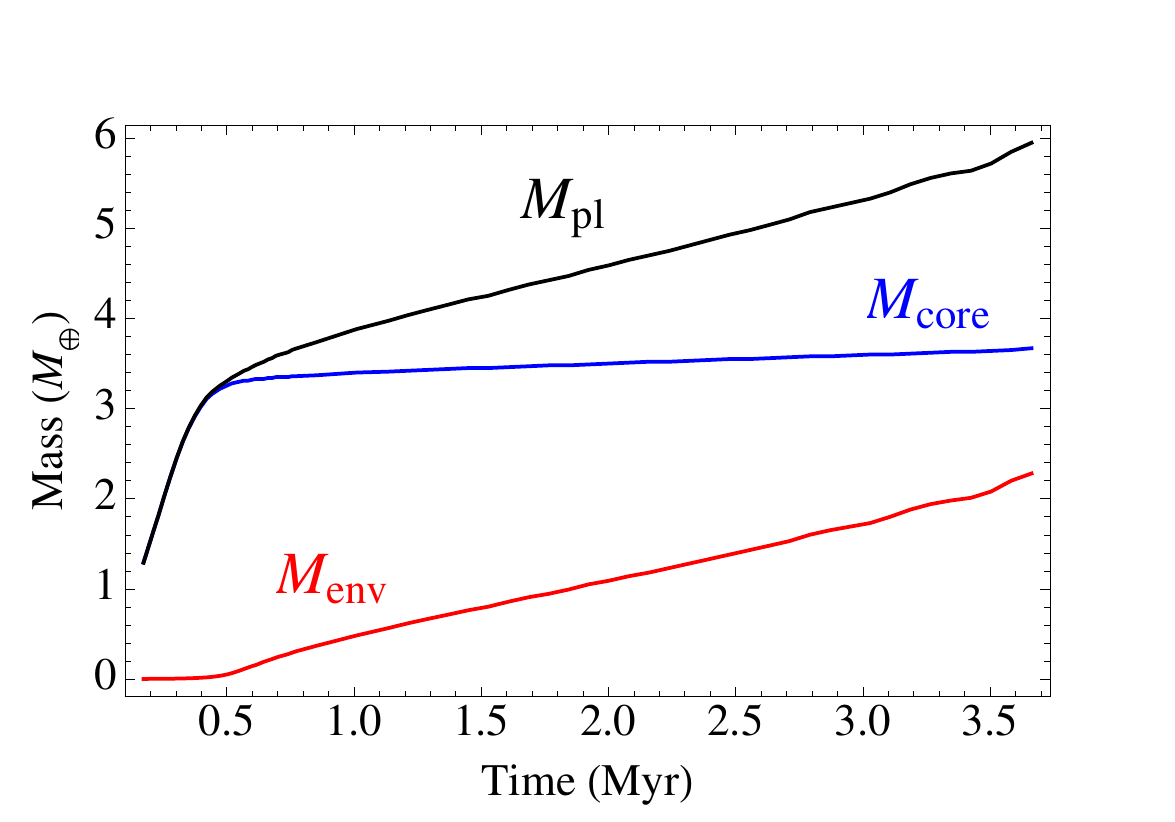}
\includegraphics[angle=0,height=6.cm]{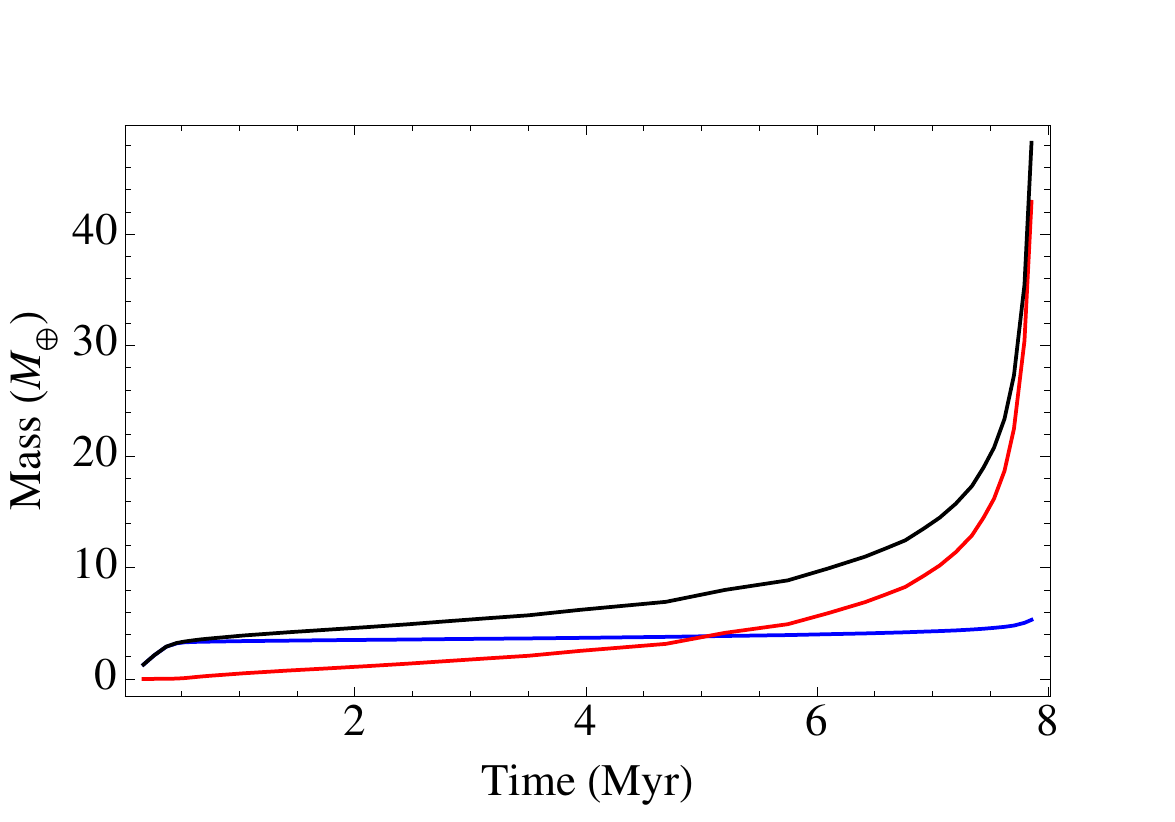}
\includegraphics[angle=0,height=6.cm]{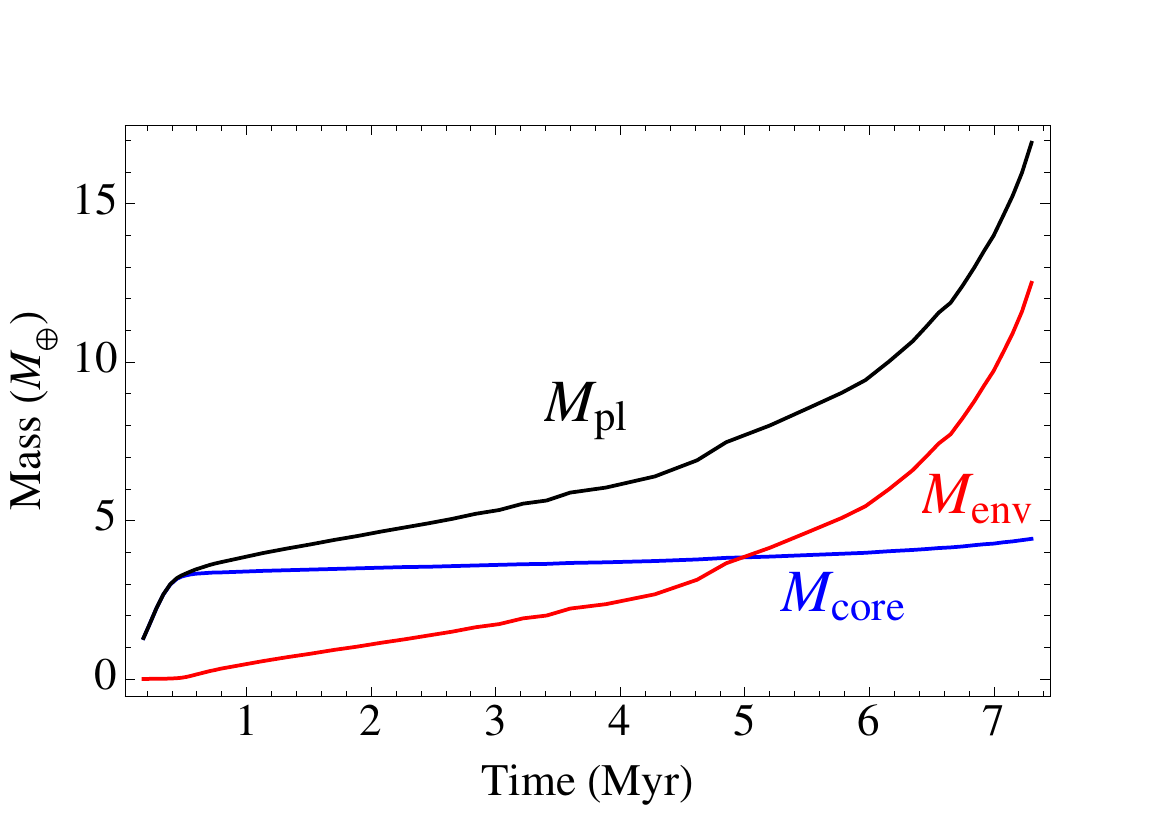}
\caption{Run 20UN1: planetary growth at 20 AU with solid surface density $\sigma_s$=0.35 
g cm$^{-2}$ and high accretion 
rate - [dM/dt]$_{HIGH}$ (Eq.~2). The planetesimal size is 1 km. The red, blue, and 
black curves represent the mass of the gaseous envelope (hydrogen and helium), core (heavy elements), and total planetary mass, respectively. 
{\bf Top}: planetary growth up to 3.7 Myr. {\bf Middle}: planetary growth 
up to 8 Myr. {\bf Bottom:} planetary growth up to $\approx$ 7.2 Myr,  
at which time the total mass is 17 M$_{\oplus}$.  
}
\end{center}
\end{figure}

\begin{figure}
\begin{center}
\includegraphics[angle=0,height=6.cm]{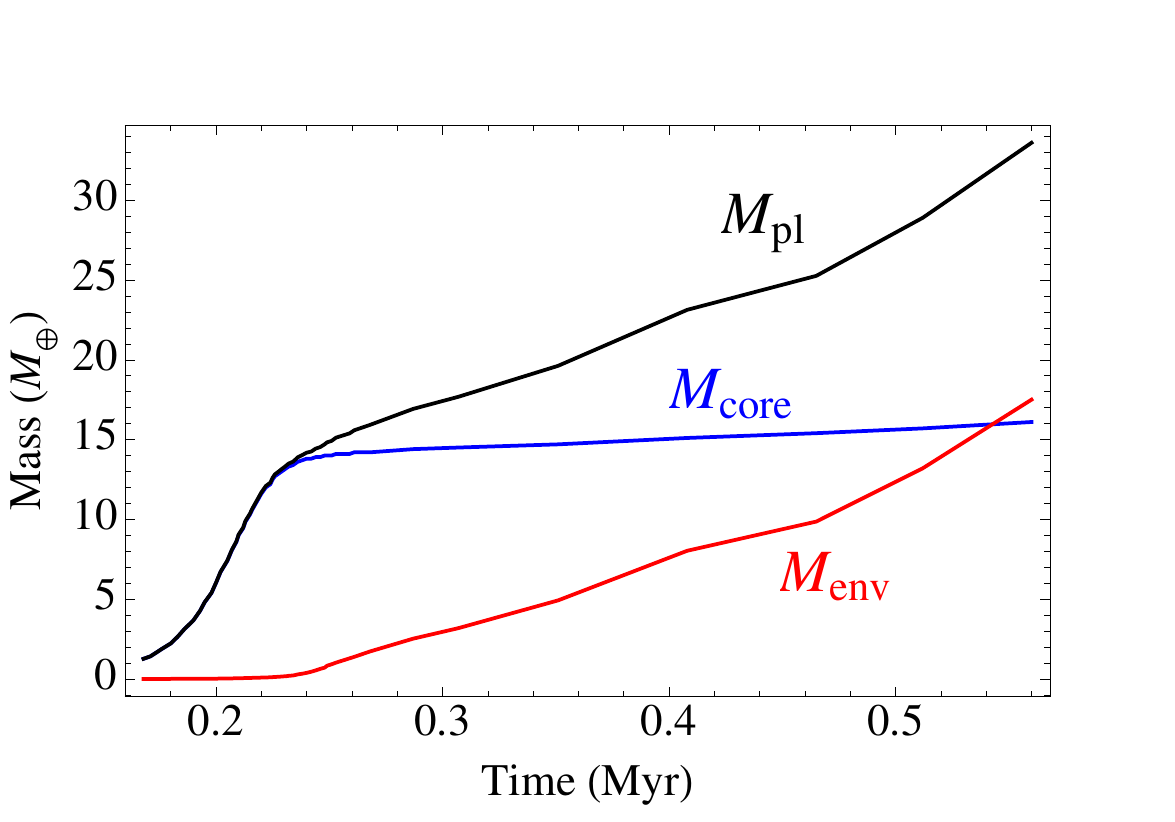}
\includegraphics[angle=0,height=6.cm]{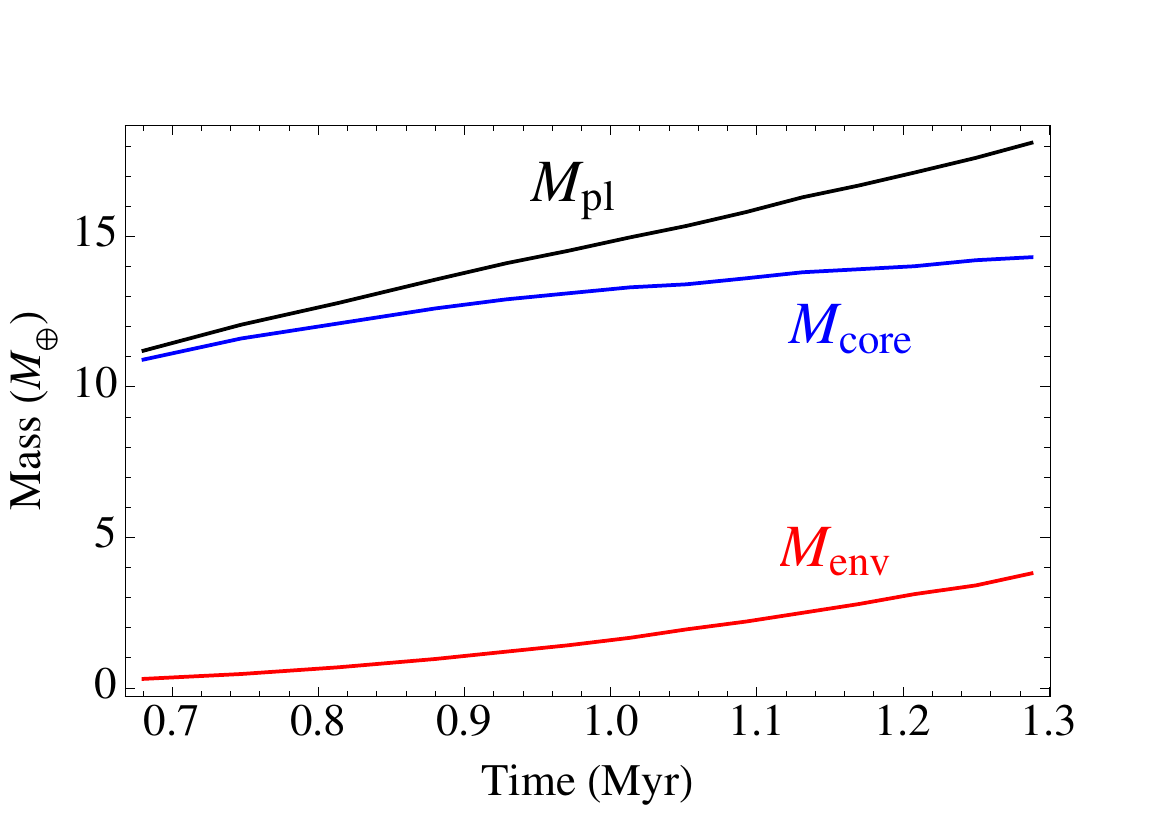}
\includegraphics[angle=0,height=6.cm]{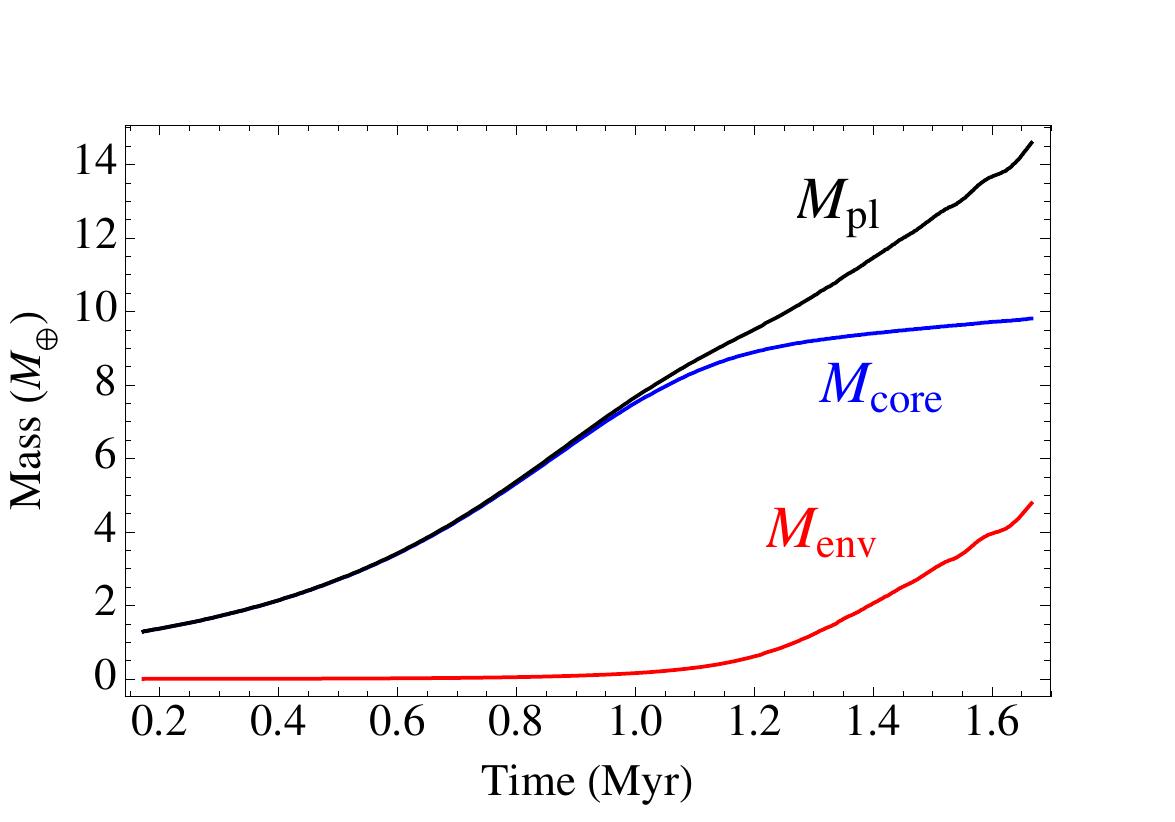}
\caption{Planetary growth at 20 AU, 
. {\bf Top:} Same as Fig.~1 but with $\sigma_s$=3.5 g cm$^{-2}$ (Run 20UN2). 
{\bf Middle:} Simulation without planetesimal scattering or envelope enhancement of
the core accretion cross section, with $\sigma_s$=0.7 g cm$^{-2}$ and with
the high core accretion rate [dM/dt]$_{HIGH}$, Eq. (2) (Run 20UN3).
 {\bf Bottom:} Simulation with  $\sigma_s$=1.7 g cm$^{-2}$ 
including planetesimal scattering and  envelope enhancement of the
cross section with 1 km-sized planetesimals.  The  core 
accretion rate is taken from Eq.~(5), i.e., the transitional accretion rate 
(Run 20UN4).}
\end{center}
\end{figure}

\begin{figure}
\begin{center}
\includegraphics[angle=0,height=8cm]{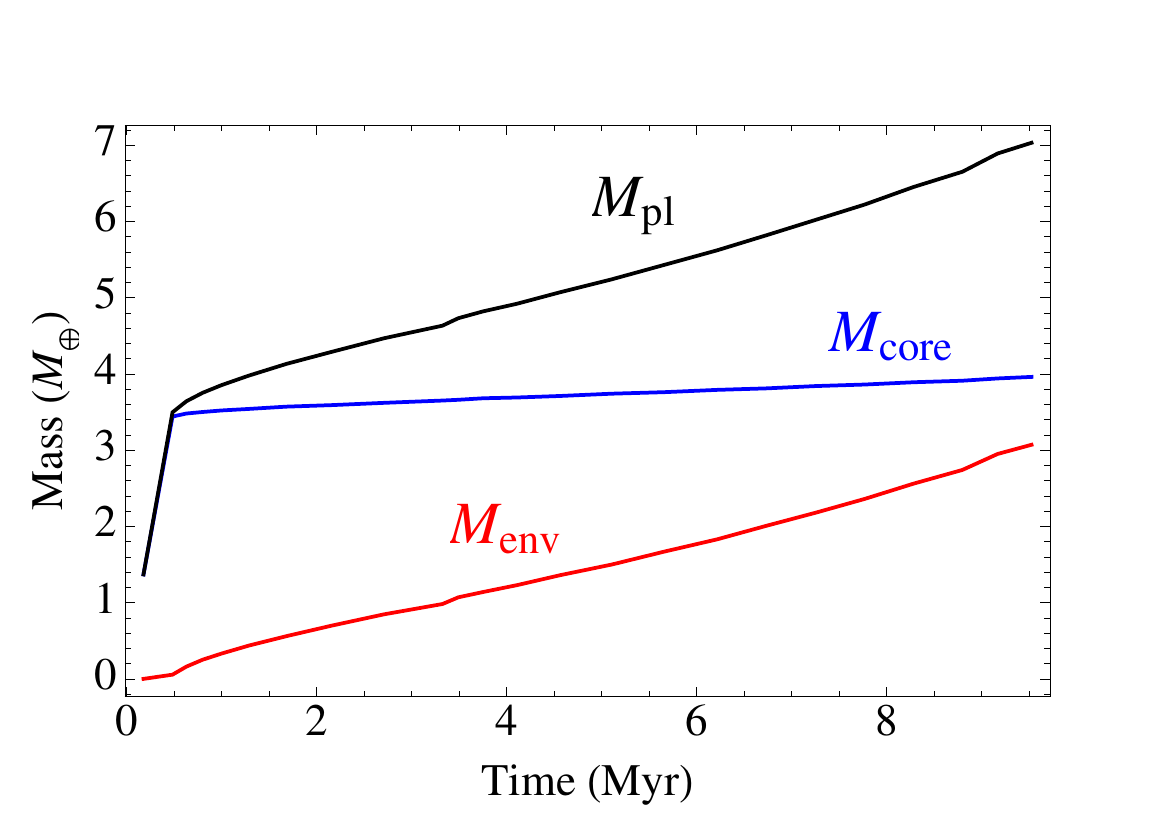}
\includegraphics[angle=0,height=8cm]{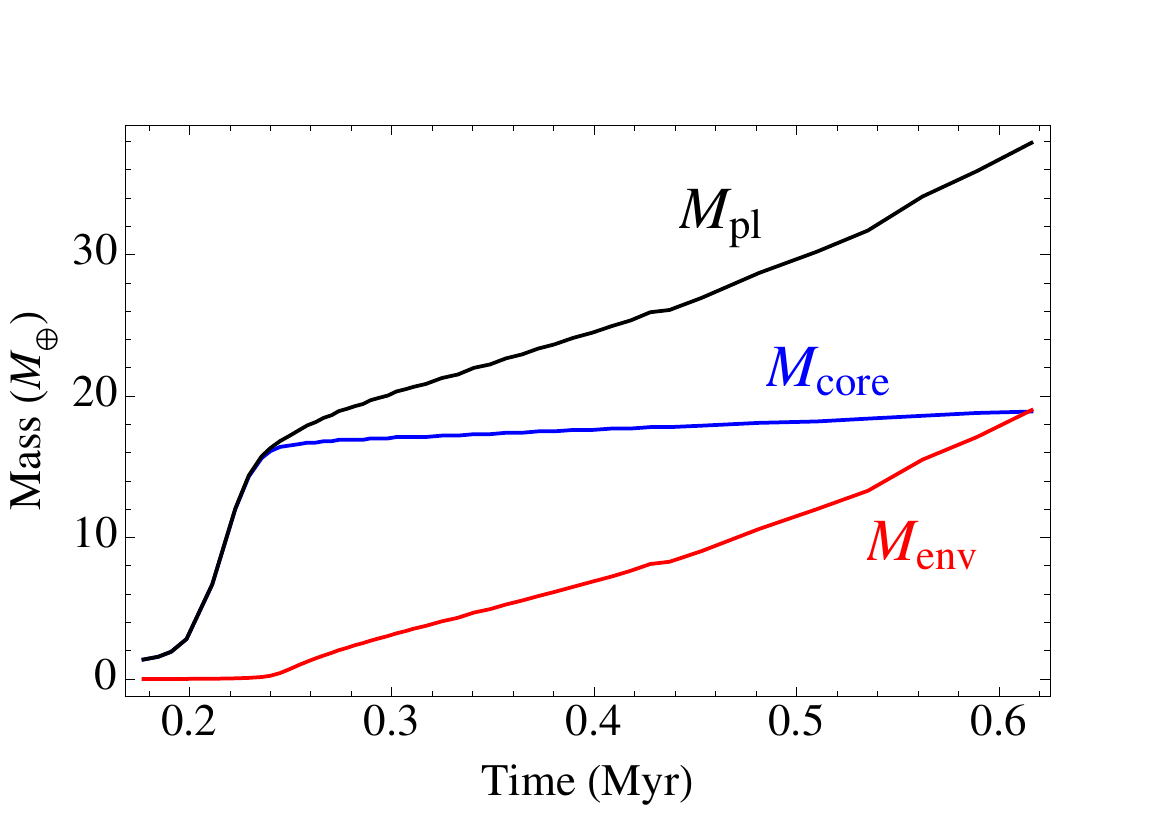}
\caption{{\bf Top:} Run 15UN1-- planetary growth at 15 AU with $\sigma_s$=0.55 g cm$^{-2}$ and 
accretion rate [dM/dt]$_{HIGH}$ (Eq.~2). {\bf Bottom:} Run 15UN2--  
planetary growth at 15 AU with $\sigma_s$=5.5 g cm$^{-2}$ and accretion rate - 
[dM/dt]$_{HIGH}$ (Eq.~2).}
\end{center}
\end{figure}

\begin{figure}
\begin{center}
\includegraphics[angle=0,height=6.1cm]{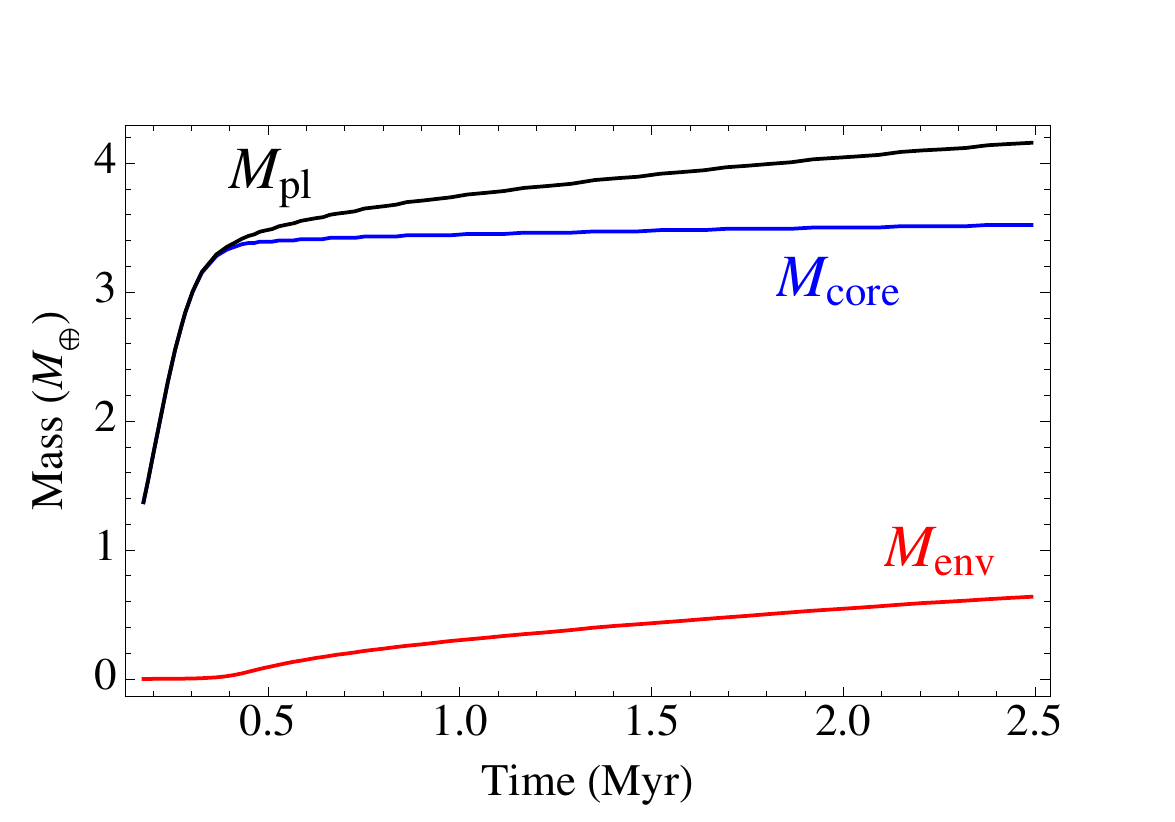}
\includegraphics[angle=0,height=6.1cm]{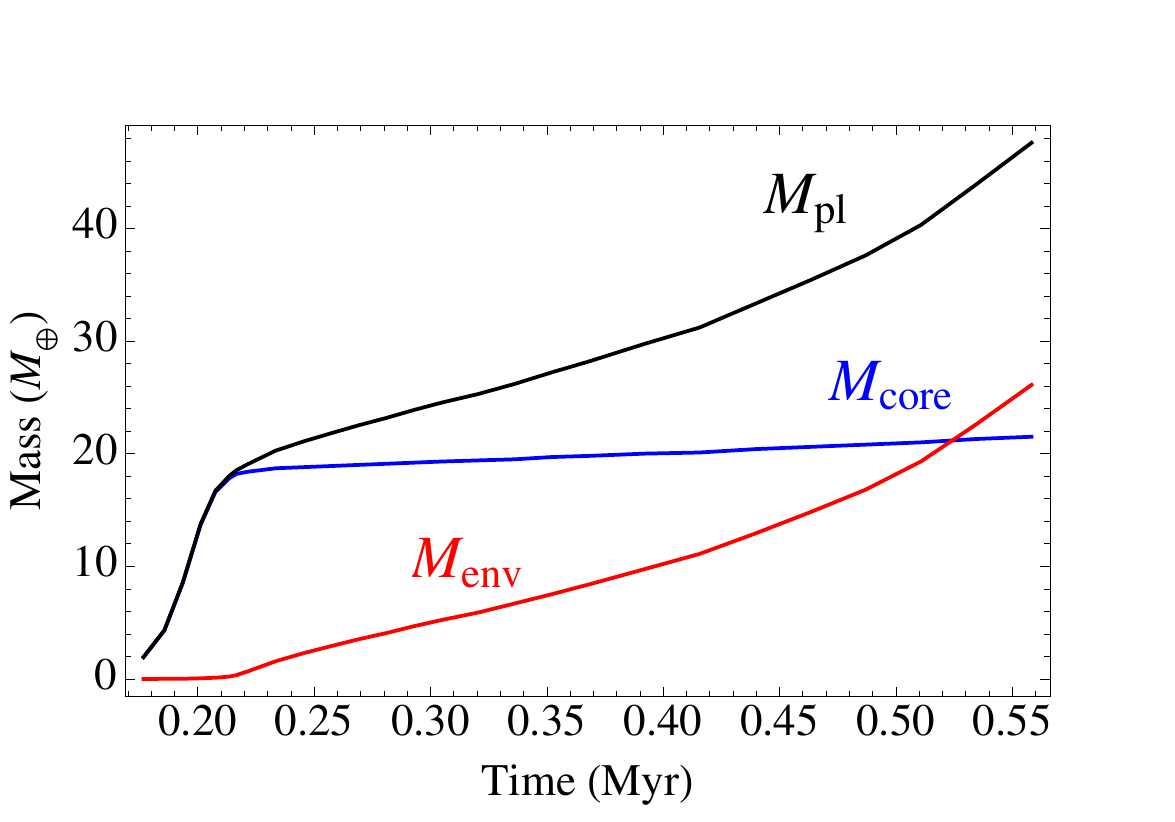}
\caption{{\bf Top:}  Run 12UN1-- Planetary growth at 12 AU with $\sigma_s$=0.75 g cm$^{-2}$ and 
the high accretion rate  [dM/dt]$_{HIGH}$ (Eq.~2). The planetesimal size 
is 1 km. The red, blue, and black curves represent the mass of the 
gaseous envelope (mostly hydrogen and helium), core (heavy elements), 
and total planetary mass, respectively. {\bf Bottom}:  Run 12UN2--   
planetary growth at 12 AU 
with $\sigma_s$=7.5 g cm$^{-2}$ and [dM/dt]$_{HIGH}$ (Eq.~2). 
}
\end{center}
\end{figure}

\begin{figure}
\begin{center}
\includegraphics[angle=0,height=8cm]{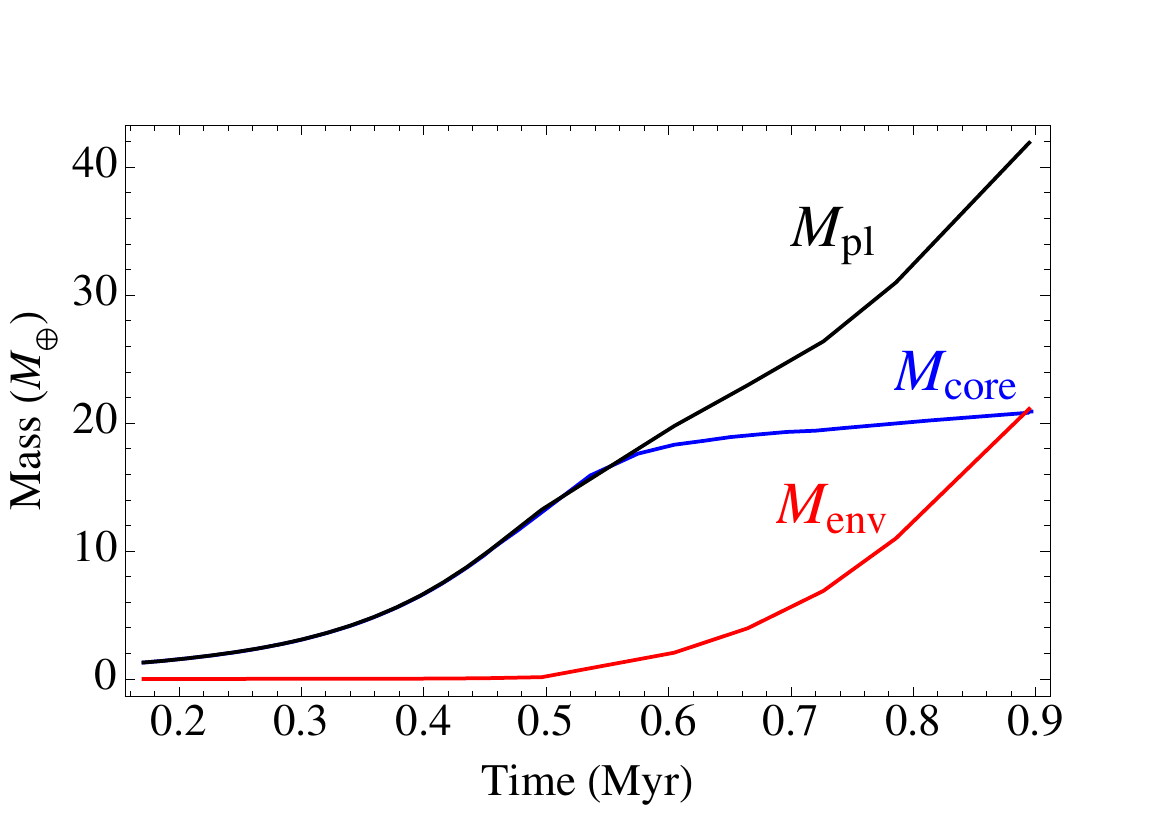}
\includegraphics[angle=0,height=8cm]{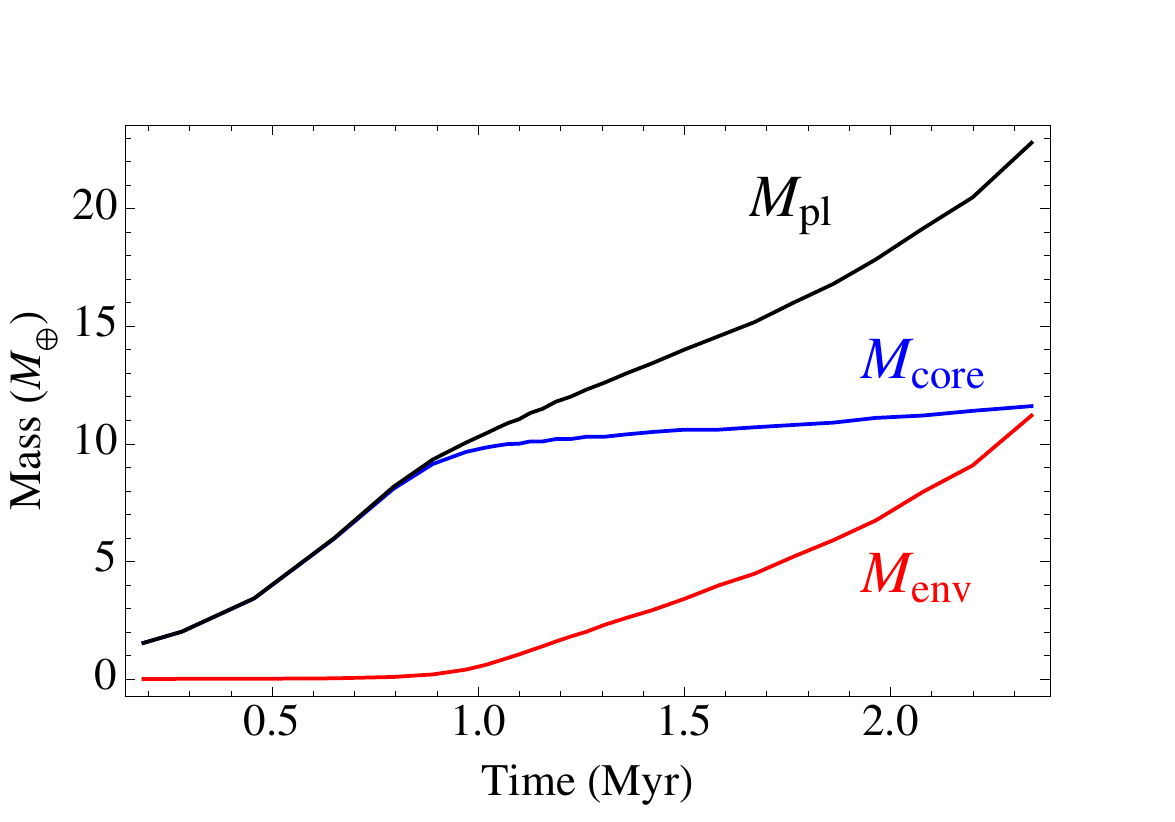}
\caption{{\bf Top:} Run 20UN3-- planetary growth at 12 AU with $\sigma_s$=7.5 g cm$^{-2}$ and the 
transitional accretion rate [dM/dt]$_{LOW}$ (Eq. 5). {\bf Bottom:} 
Run 20UN4--planetary growth at 12 AU with $\sigma_s$=3 g cm$^{-2}$ and the 
transitional accretion rate [dM/dt]$_{LOW}$ (Eq. 5).}
\end{center}
\end{figure}


\begin{figure}
\begin{center}
\includegraphics[angle=0,height=8cm]{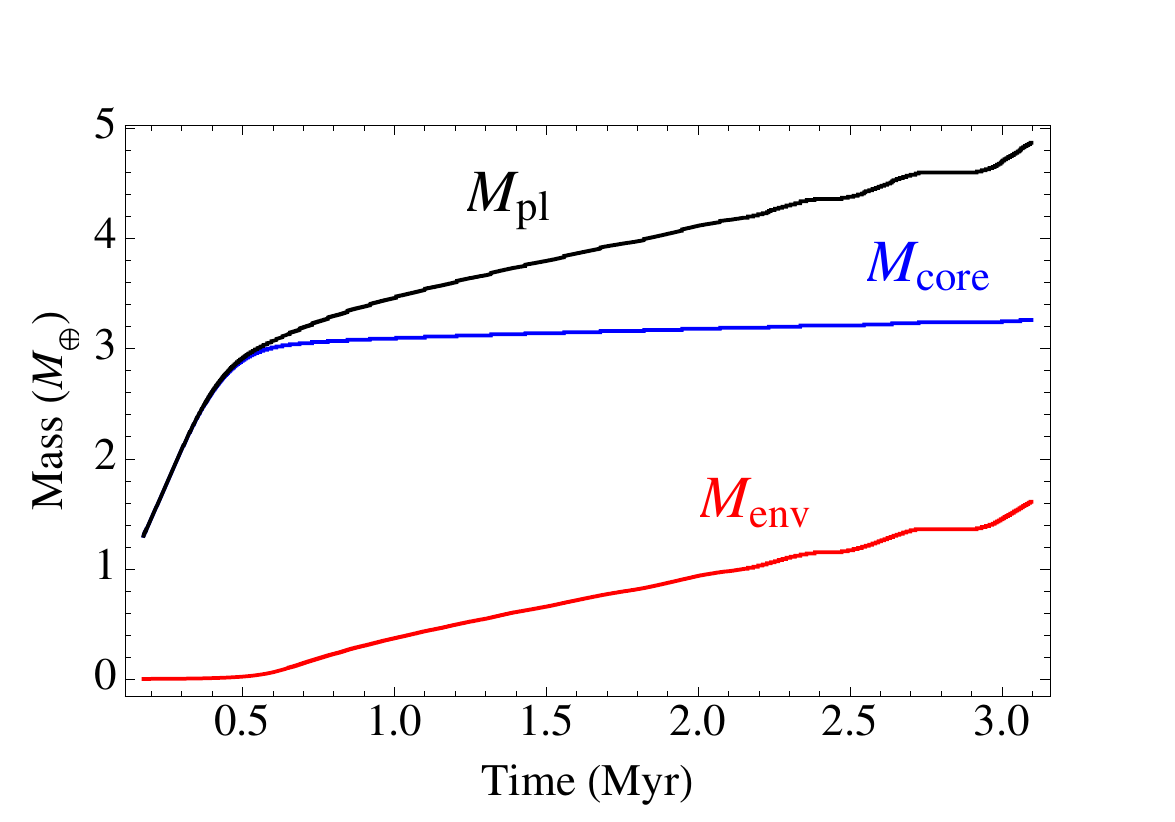}
\includegraphics[angle=0,height=8cm]{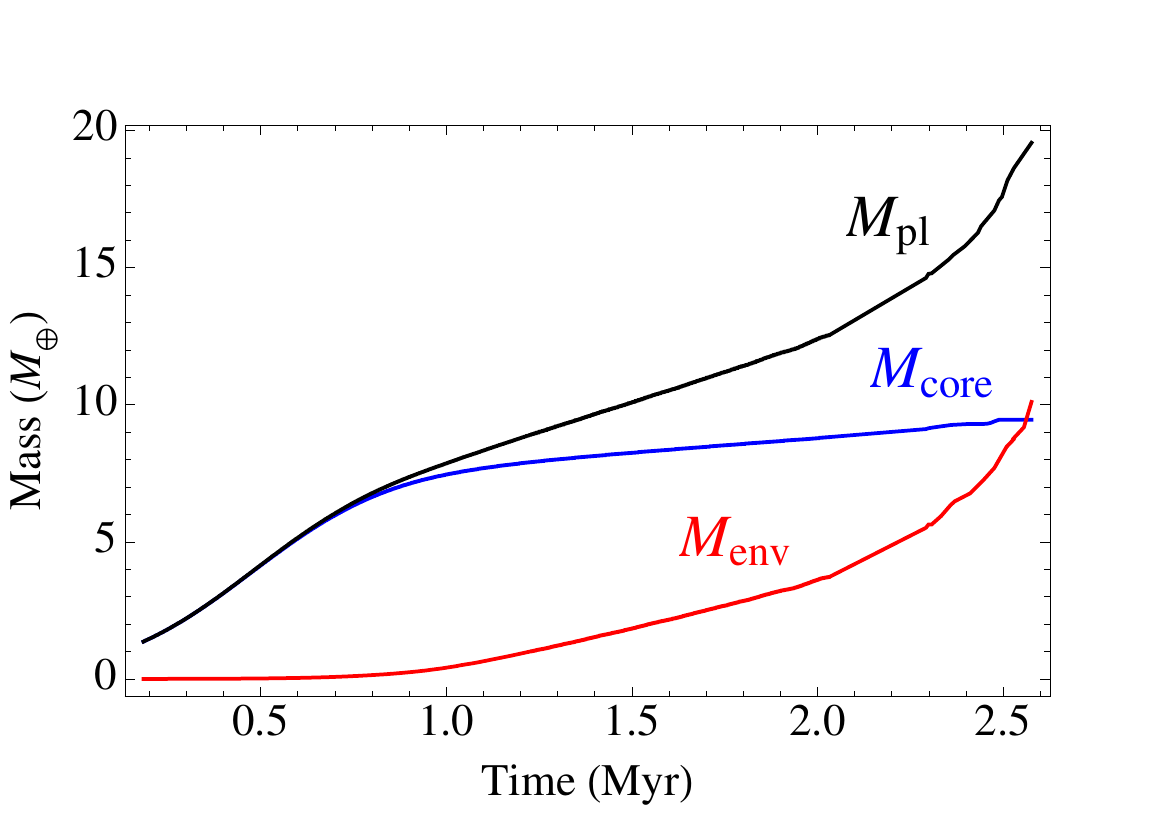}
\caption{{\bf Top:} Run 30UN1-- Planet formation at 30 AU assuming $\sigma_s$=0.2 
g cm$^{-2}$ and [dM/dt]$_{HIGH}$ (Eq.~2). {\bf Bottom:} Run 30UN2--Same as top panel but without planetesimal scattering. }
\end{center}
\end{figure}

\begin{figure}
\begin{center}
\includegraphics[angle=0,height=8cm]{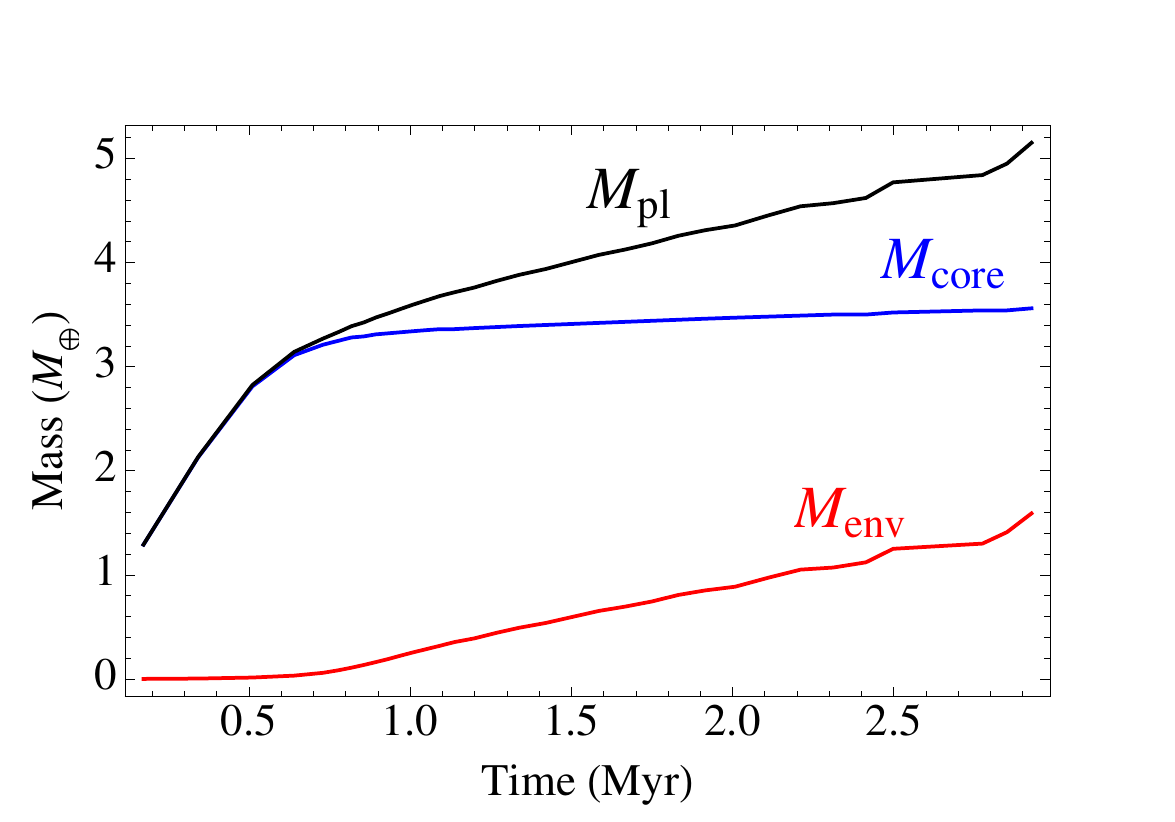}
\includegraphics[angle=0,height=8cm]{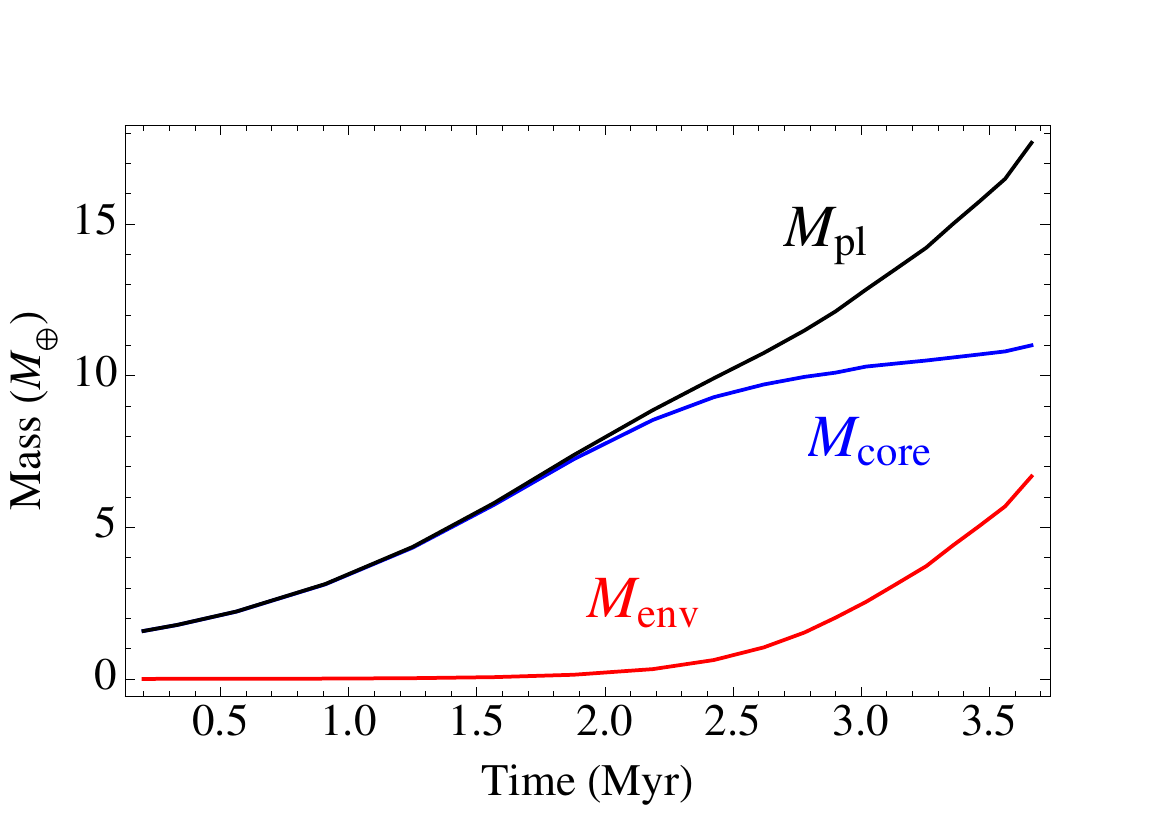}
\caption{ Planetary growth with 100 km-sized planetesimals. 
{\bf Top:} Run 20UN5-- planetary formation at 20 AU, 
$\sigma_s$=0.35 g cm$^{-2}$ 
with core accretion rate [dM/dt]$_{HIGH}$, i.e. Eq. (2). {\bf Bottom:} Run 
12UN5--  
planetary formation at 12 AU with $\sigma_s$=3 g cm$^{-2}$ and the low 
core accretion rate [dM/dt]$_{LOW}$ (Eq. 5).}
\end{center}
\end{figure}

\end{document}